\documentclass[preprint,float]{JHEP3}
\usepackage{epsfig}
\def\@spires#1{\href{http://www-spires.slac.stanford.edu/spires/find/hep/www?j=#1}} 
\newcommand\jcap[3] {\@spires{JCAP
		{{\it J.\ Cosm.\ Astro.\ Phys.\ }{\bf #1} (#2) #3}}}
\def\eslt{\not\!\!{E_T}}

\def\to{\rightarrow}

\def\bi{\begin{itemize}}
\def\ei{\end{itemize}}
\def\be{\begin{equation}}
\def\ee{\end{equation}}
\def\bea{\begin{eqnarray}}
\def\eea{\end{eqnarray}}
\def\te{\tilde e}

\def\tu{\tilde u}

\def\tb{\tilde b}
\def\tc{\tilde c}

\def\td{\tilde d}

\def\tst{\tilde t}
\def\ttau{\tilde \tau}
\def\tmu{\tilde \mu}
\def\tg{\tilde g}
\def\tnu{\tilde\nu}
\def\tell{\tilde\ell}
\def\tq{\tilde q}
\def\tw{\widetilde W}
\def\tz{\widetilde Z}
\def\alt{\stackrel{<}{\sim}}
\def\agt{\stackrel{>}{\sim}}
\title{
Direct, Indirect and Collider Detection\\ of Neutralino Dark Matter
In SUSY Models\\ with Non-universal Higgs Masses
}

\author{Howard Baer, Azar Mustafayev, Stefano Profumo,
\\ Department of Physics, Florida State University\\ 
Tallahassee, FL 32306, USA\\
E-mail: \email{baer@hep.fsu.edu},
\email{mazar@hep.fsu.edu},\email{profumo@hep.fsu.edu}}
\author{Alexander Belyaev,
\\ Department of Physics and Astronomy, Michigan State University\\ 
East Lansing, MI 48824, USA\\
E-mail: \email{belyaev@pa.msu.edu}}
\author{Xerxes Tata
\\ Department of Physics and Astronomy, University of Hawaii,\\
Honolulu, HI 96822, USA \\ 
E-mail: \email{tata@phys.hawaii.edu}}

\preprint{\vbox{\hbox{FSU-HEP-050315} \vspace{0.2cm}
                 \hbox{UH-511-1067-05}}} 

\abstract{
In supersymmetric models with gravity-mediated SUSY breaking,
universality of soft SUSY breaking sfermion masses $m_0$ is motivated
by the need to suppress unwanted flavor changing 
processes.
The same motivation, however, does not apply to soft breaking Higgs masses,
which may in general have independent masses from matter scalars at the
GUT scale. We explore phenomenological implications of both the one-parameter
and two-parameter non-universal Higgs mass
models (NUHM1 and NUHM2), and examine the parameter ranges compatible 
with $\Omega_{CDM}h^2$, $BF(b\to s\gamma )$ and
$(g-2)_\mu$ constraints. 
In contrast to the mSUGRA model, 
in both NUHM1 and NUHM2 models, the 
dark matter $A$-annihilation funnel can be reached at 
low values of $\tan\beta$, while the higgsino dark matter 
annihilation regions can be reached for low values of $m_0$.
We show that there may be observable
rates for indirect and direct detection of neutralino cold dark matter
in phenomenologically aceptable ranges of parameter space.
We also examine implications of the NUHM models for the Fermilab Tevatron,
the CERN LHC and a $\sqrt{s}=0.5-1$ TeV $e^+e^-$ linear collider.
Novel possibilities include:
very light $\tu_R,\ \tc_R$ squark and $\te_L$ slepton masses as well as 
light charginos and neutralinos and $H,\ A$ and $H^\pm$ 
Higgs bosons. 
}


\begin{document}

\section{Introduction}
\label{sec:intro}
The minimal supergravity (mSUGRA) model~\cite{msugra} provides a
convenient and popular template for exploration of many of the
phenomenological consequences of weak scale supersymmetry~\cite{rev}.  In
mSUGRA, it is assumed that supersymmetry is broken in a hidden
sector of the model, with SUSY breaking communicated to the visible
sector via gravitational interactions. The qualifier ``minimal'' in
mSUGRA refers to the assumption of a flat K\"ahler metric, which leads
to universal tree level scalar masses at some high energy scale, usually
taken to be $Q=M_{GUT}$. The universality assumption ensures the
super-GIM mechanism~\cite{dg}, which suppresses unwanted flavor-changing
neutral current effects.  An attractive feature of this framework is
that electroweak symmetry can be radiatively broken (REWSB). 
This allows one to
eliminate the superpotential $|\mu |$ parameter in favor of $M_Z$, and
the low energy phenomenology is then determined by the well-known
parameter space
\begin{equation}
{\rm mSUGRA}:\ \ \ m_0,\ m_{1/2},\ A_0,\ \tan\beta ,\ {\rm and}\ sign(\mu ) .
\end{equation}
Here $m_0$ is the common GUT scale scalar mass, $m_{1/2}$ is the
common GUT scale gaugino mass, $A_0$ is the common GUT scale trilinear
term, $\tan\beta$ is the weak scale ratio of Higgs field vacuum
expectation values, and $\mu$
is the superpotential Higgs mass term. We take 
$m_t=178$ GeV throughout this paper.

The mSUGRA model has been criticized because the assumption
of universal scalar masses is {\it ad hoc} and does 
not follow from any known symmetry principle~\cite{sug_crit}.
While it is possible to invoke an additional
global $U(N)$ symmetry for the (gravitational)
interactions of the $N$ chiral supermultiplets, this symmetry
is clearly not respected by superpotential Yukawa couplings, 
and  
radiative corrections involving
these Yukawa interactions can lead to large
deviations from the universality hypothesis~\cite{bagger}. 

The assumption of equality of scalar masses receives partial
support in Grand Unified Theories (GUTs). For instance, in $SO(10)$
SUSY GUT models, all matter superfields of a single generation belong to
a 16 dimensional spinor representation $\hat{\psi}(16)$ of $SO(10)$,
and their mass degeneracy is guaranteed if SUSY
breaking masses are acquired 
above the $SO(10)$ breaking scale. If the mechanism by
which matter scalars acquire SUSY breaking masses is generation blind, 
universality of matter scalar masses would then obtain. 
In the case of minimal $SO(10)$ SUSY GUTs, the two MSSM Higgs doublet
superfields $\hat{H}_u$ and $\hat{H}_d$ belong to the same 10 dimensional
fundamental representation $\hat{\phi}(10)$, so the corresponding SUSY
breaking scalar mass terms would not be expected to be
the same as those of the
matter scalars.
The phenomenologically 
desirable super-GIM mechanism would be ensured by requiring 
a $U(3)$ symmetry  
amongst the different generations. In practice, the
amount of degeneracy needed is greatest for the first
two generations where
FCNC constraints are the strongest,
while  the corresponding constraints for the third generation
are rather mild~\cite{nmh}. 
The need for generational
degeneracy can be further reduced if one invokes as well a degree of alignment
between squark and quark mass matrices, or a (partial) decoupling
solution to the SUSY flavor problem.

In this paper, we will maintain degeneracy amongst matter scalars
at scales $Q\simeq M_{GUT}$, but will allow non-universality to
enter the model via soft SUSY breaking masses for the Higgs scalars.
In our analysis, we will differentiate between two cases for the 
non-universal Higgs mass (NUHM)
models. Inspired by GUT models where both MSSM Higgs doublets are
contained in a single superfield, we will first examine the NUHM model
where $m_{H_u}^2=m_{H_d}^2\ne m_0^2$ at $Q=M_{GUT}$.  In this case, we
define the new parameter $m_\phi
=sign(m_{H_u,d}^2)\cdot\sqrt{|m_{H_{u,d}}^2|}$ at the GUT scale. Thus,
the parameter space of this one parameter extension of the mSUGRA model
is given by,
\begin{equation}
{\rm NUHM1}:\ \ \ 
m_0,\ m_\phi ,\ m_{1/2},\ A_0,\ \tan\beta \ {\rm and}\ sign(\mu ).
\end{equation}

The second case 
is inspired by GUT models where $\hat{H}_u$ 
and $\hat{H}_d$ belong to different multiplets.
The parameter space for this second case is then given by
\begin{equation}
{\rm NUHM2}:\ \ \ 
m_0,\ m_{H_u}^2,\ m_{H_d}^2,\ m_{1/2},\ A_0,\ \tan\beta \ {\rm and}\ 
sign(\mu ) .
\end{equation}
The conditions of electroweak gauge symmetry breaking allows one to 
trade the GUT scale masses $m_{H_u}^2$ and $m_{H_d}^2$ 
for the weak scale parameters $\mu$ and $m_A$. 

We remark that regardless of any theoretical motivation, if any small
departure from a well-motivated framework such as mSUGRA causes
significant differences in the phenomenological outcome, the new
framework is worthy of examination. We will see below that enlarging the
model parameter space to split off the GUT scale Higgs boson mass
parameters from those of other scalars leads to significant departures
from mSUGRA expectations upon the incorporation of the WMAP constraint
on the relic density of cold dark matter.  Motivated by this, our goal
here is to explore in detail the phenomenological consequences of the
NUHM1 and NUHM2 models.  Before doing so, we note that the mSUGRA model
has recently been tightly constrained by several
measurements~\cite{bbbmtw}. These include 1.) the combined
measurement~\cite{bsg} of the branching fraction $BF(b\to s\gamma
)=(3.25\pm 0.37)\times 10^{-4}$, 2.) the measurement~\cite{e821} of the
deviation of the muon anomalous magnetic moment $\Delta a_\mu\equiv
\Delta (g-2)_\mu/2 =(27\pm 10)\times 10^{-10}$ from the SM
prediction~\cite{gm2_sm}, and 3.) the WMAP determination of the relic
density of cold dark matter (CDM) in the universe~\cite{wmap}:
$\Omega_{CDM}h^2= 0.113\pm 0.009$.  In addition, we invoke the usual
constraint from LEP2 that charginos should have mass $m_{\tw_1}\ge
103.5$ GeV.  We remark that there could be significant theoretical
uncertainties in the evaluation of both $\Delta a_\mu$ and, especially
for large
values of $\tan\beta$, also $BF(b\to s\gamma)$ so that any inferences
from them should be interpreted with care.  The first two constraints
favor mSUGRA models with $\mu >0$.  The WMAP constraint restricts the
mSUGRA parameter space to lie in one of the following
regions~\cite{ellis_wmap,our_relic,csaba,nanop}:
\begin{itemize} 
\item the bulk region at low $m_0$ and low $m_{1/2}$, where
neutralino annihilation in the early universe occurs predominantly via
$t$-channel slepton exchange (this region is now essentially excluded by the
combination of WMAP $\Omega_{CDM}h^2$ bound and the LEP2 bounds on
$m_{\tw_1}$ and $m_h$, save where it overlaps with the stau 
co-annihilation region),
\item the stau co-annihilation at low $m_0$ but almost any $m_{1/2}$ value, 
where $m_{\tz_1}\simeq m_{\ttau_1}$~\cite{stau}, or the stop
co-annihilation region for special values of $A_0$ where $m_{\tz_1}\simeq
m_{\tst_1}$~\cite{stop}, 
\item the hyperbolic branch/focus point region (HB/FP) at large $m_0$, 
where $|\mu |$ becomes small, and the neutralino develops a significant
higgsino component~\cite{ccn,fmm,bcpt_lhc}, and
\item the $A$-annihilation funnel at large $\tan\beta$, where
$2m_{\tz_1}\sim m_A$, and neutralino annihilation in the early universe
occurs via the broad $A$ and $H$ Higgs boson resonances~\cite{Afunnel}.
A light Higgs resonance annihilation region may also be possible
at low $m_{1/2}$ values where $2m_{\tz_1}\simeq m_h$~\cite{lighth}.

\end{itemize}
Throughout this work, we use Isajet 7.72 to generate sparticle mass
spectra~\cite{isajet}, IsaReD~\cite{our_relic} for 
the relic density calculation, and the {\tt DarkSUSY} 
package~\cite{Gondolo:2004sc} for the computation of 
dark matter detection rates.

At this point, we would like to take note of a variety of earlier studies
of models with non-universal Higgs masses. SUGRA models with non-universal
Higgs masses were first studied by Berezinski {\it et al.}, who focussed on
direct detection of neutralino dark matter~\cite{berez1} and indirect
detection via neutrino telescopes\cite{berez2}. Around the same time,
direct detection of neutralino dark matter in NUHM models was also
investigated by Arnowitt and Nath~\cite{an}, and subsequently by Bottino
{\it et al.}\cite{bott1}. These latter papers explored only the case of
positive squared Higgs masses.  Bottino {\it et al.} explored direct DM
detection for cases including negative squared Higgs masses in Ref.
\cite{bott2}. Ellis {\it et al.}~\cite{efos1} made a thorough exploration
of parameter space of the NUHM2 model using the $\mu$ and $m_A$ variables,
and investigated direct detection rates in Ref. \cite{efos2}. Indirect
detection via neutrinos was investigated by Barger {\it et
al.}\cite{vernon} for models with universality and non-universality.
Recently, both direct\cite{munoz1} and indirect\cite{munoz2} detection of
neutralino dark matter has been investigated by Munoz {\it et al.} in the
context of models with {\it both} scalar and gaugino mass
non-universality.  

Our present study goes beyond these previous works in several respects:\\
1)~we investigate the more constrained NUHM1 model in Sec. \ref{sec:nuhm1},
and show for the first time that  in this minimal (one parametric
extension)
of mSUGRA model there are always {\it two solutions for low relic
density}:
one is  neutralino annihilation via heavy Higgs resonance even at low
values
of $\tan\beta$ while the other is neutralino annihilation via higgsino
components
at low values of $m_0$;\\
2)~we find new allowed regions of the NUHM2 model
-- the light squark/slepton co-annihilation regions
as discussed
in Sec. \ref{sec:nuhm2};\\
3)~we investigate
direct and indirect detection of neutralino dark matter, including
anti-matter, neutrino and gamma ray indirect searches
for the new parameter regions mentioned above;\\
4)~for the first time, we consider the implications of the NUHM1 and NUHM2
models for collider
searches at the Fermilab Tevatron, CERN LHC and ILC linear $e^+e^-$
colliders, and show 
how these correlate with the direct and indirect dark matter
searches. In this connection, we also emphasize the sensitivity of the
implications of the WMAP measurement of $\Omega_{CDM}h^2$ for collider
expectations to the underlying framework.  In particular, we show that
inferences valid in the mSUGRA model may simply be invalid in the extended
NUHM framework.

The remainder
of this paper is organized as follows. 
In Sec. \ref{sec:nuhm1}, we explore the allowed regions of the NUHM1
model which was first studied in Ref.~\cite{bbmpt}.
We will find that even for low values of $m_0$, 
{\it raising} the ratio $m_\phi /m_0$ brings us into the low $|\mu |$
region where the
relic density is in accord with the WMAP allowed range; this is
quite unlike the situation in mSUGRA where the higgsino annihilation
region occurs only at multi-TeV values of $m_0$.  In addition, 
{\it lowering} the ratio $m_\phi /m_0$ into the range of {\it negative} 
values decreases the value of $m_A$ until the $A$-annihilation
funnel is reached. In this case, the $A$-funnel region
can occur at {\it any} $\tan\beta$ value where an acceptable spectrum can
be generated. 

We introduce and outline in Sec~\ref{sec:dm} the
computation of direct and indirect dark matter rates. We make use
of consistent halo models in the attempt to systematically
compare the reach in all various detection channels.
We find enhanced signal rates for direct and indirect
detection of neutralino cold dark matter in these 
WMAP-allowed regions~\cite{bo}.

In Sec. \ref{sec:nuhm1_col}, we explore some unique consequences of the
NUHM1 model for collider searches.
In the higgsino region of the NUHM1 model, 
charginos and neutralinos all become rather light,
and more easily accessible to collider searches. In addition, 
lengthy gluino and squark cascade decays to the various charginos 
and neutralinos occur, leading to the possibility of 
spectacular events at the CERN LHC. In the $A$-funnel region, the
$A,\ H$ and $H^\pm$ Higgs bosons may be kinematically accessible to 
searches at the International Linear Collider (ILC) or at the CERN LHC,
and may also be present in gluino and squark cascade decays.

In Sec. \ref{sec:nuhm2}, we explore the NUHM2 model. In this case, since
$\mu$ and $m_A$ can now be used as input parameters, it is always 
possible to choose values such that one lies either in the higgsino
annihilation region or in the $A$-funnel region, for
any value of $\tan\beta$, $m_0$ or $m_{1/2}$ that gives rise to a 
calculable SUSY mass spectrum. In the low $\mu$ region, charginos and 
neutralinos are again likely to be light, and possibly accessible to
Fermilab Tevatron, CERN LHC and ILC searches. 
If instead one is in the $A$-annihilation funnel, 
then the heavier Higgs scalars may be light enough 
to be produced at observable rates at hadron or lepton colliders. 
In addition, new regions are found where consistency with WMAP data is obtained
because either $\tu_R,\ \tc_R$ squarks or left- sleptons become very light.
The $\tu_R$ and $\tc_R$ co-annihilation region leads to large rates for
direct and indirect detection of neutralino dark matter, and is in fact
already constrained by searches from CDMS2.
We present a summary and our conclusions in Sec.~\ref{sec:conclusions}.

\section{NUHM1 model}
\label{sec:nuhm1}

\subsection{Overview}

In this section, we investigate the phenomenology of the NUHM1 model,
wherein the Higgs masses $m_{H_u}^2=m_{H_d}^2\equiv sign(m_\phi)\cdot
|m_\phi|^2$ at $Q=M_{GUT}$, with $m_\phi^2\ne m_0^2$. We
first note that the parameter range for $m_\phi$ need not be limited to
positive values at the GUT scale, and that, indeed, to achieve radiative
EWSB, $m_{H_u}^2$ must evolve to negative values.  Indeed, negative
squared Higgs mass parameters (at $Q=M_{GUT}$) 
are predicted in the $SU(5)$ fixed point scenario
of Ref.~\cite{javier}.  

We also note that some authors impose so-called GUT stability (GS)
bounds,
\begin{eqnarray}
m_{H_u}^2(M_{GUT})+\mu^2(M_{GUT}) &>&0\ \ {\rm and}\\
m_{H_d}^2(M_{GUT})+\mu^2(M_{GUT}) &>&0\;,
\end{eqnarray}
to avoid
EWSB at too high a scale. The reliability of these bounds is debated in 
Refs.~\cite{gutstab}; here we will merely remark on regions of parameter
space where they occur, and leave it to the reader to decide whether or
not to impose them.

We show one of the critical aspects of the NUHM1 model in
Fig.~\ref{fig:mumavsmH}, where we plot the values of $\mu$, $m_A$,
$m_{\tz_1}$ and $2m_{\tz_1}$ versus $m_\phi$ while fixing
$m_0=m_{1/2}=300$ GeV, with $A_0=0$, $\tan\beta =10$ and $\mu >0$.  The
region to the left of the dot-dashed line indicates where the GS bound
fails.  The curves terminate because electroweak symmetry breaking is
not obtained as marked on the figure: in fact, on the right, where
$|\mu|$ becomes small, the chargino mass falls below the LEP bound
just before the EWSB constraint kicks in.  The black curve denotes the value
of the $\mu$ parameter, which takes a value of $\mu =409$ GeV for
$m_\phi=300$ GeV (the mSUGRA case).  The parameter $\mu$ becomes much
larger for $m_\phi < -m_0$, and much smaller for
$m_\phi>m_0$~\cite{drees_susy2004}.  The region of small $\mu$ is of
particular interest since in that case the lightest neutralino develops
substantial higgsino components, and leads to a relic density which can
be in accord with the WMAP determination.  In contrast, in the mSUGRA
model the higgsino-LSP region occurs in the HB/FP region, which occurs
at very large $m_0$ values of order several TeV (depending somewhat on
the assumed value of $m_t$).  The HB/FP region has been criticized in
the literature in that the large $m_0$ values may lead to large
fine-tunings~\cite{finetune} (for an alternative point of view, see
Refs.~\cite{ccn} and~\cite{fmm}).  This illustrates an important virtue
of the NUHM1 model: the higgsino annihilation region may be reached even
with {\em arbitrarily low} values of $m_0$ and $m_{1/2}$, provided of
course that sparticle search bounds are respected.

We also see from Fig.~\ref{fig:mumavsmH} that the value of $m_A$ can range
beyond its mSUGRA value for large values of $m_\phi$, to quite small
values when $m_\phi$ becomes less than zero. In particular, when
$m_A\sim 2m_{\tz_1}$, neutralinos in the early universe may
annihilate efficiently through the $A$ and $H$ Higgs resonances, so that again
$\Omega_{CDM} h^2$ may be brought into accord with the WMAP result.
In the mSUGRA model, the $A$-annihilation funnel occurs only at large
$\tan\beta\sim 45-55$. However, in the NUHM1 model, the $A$-funnel region
may be reached even for low $\tan\beta$ values if $m_\phi$ is taken to 
be sufficiently negative. 
\FIGURE{\epsfig{file=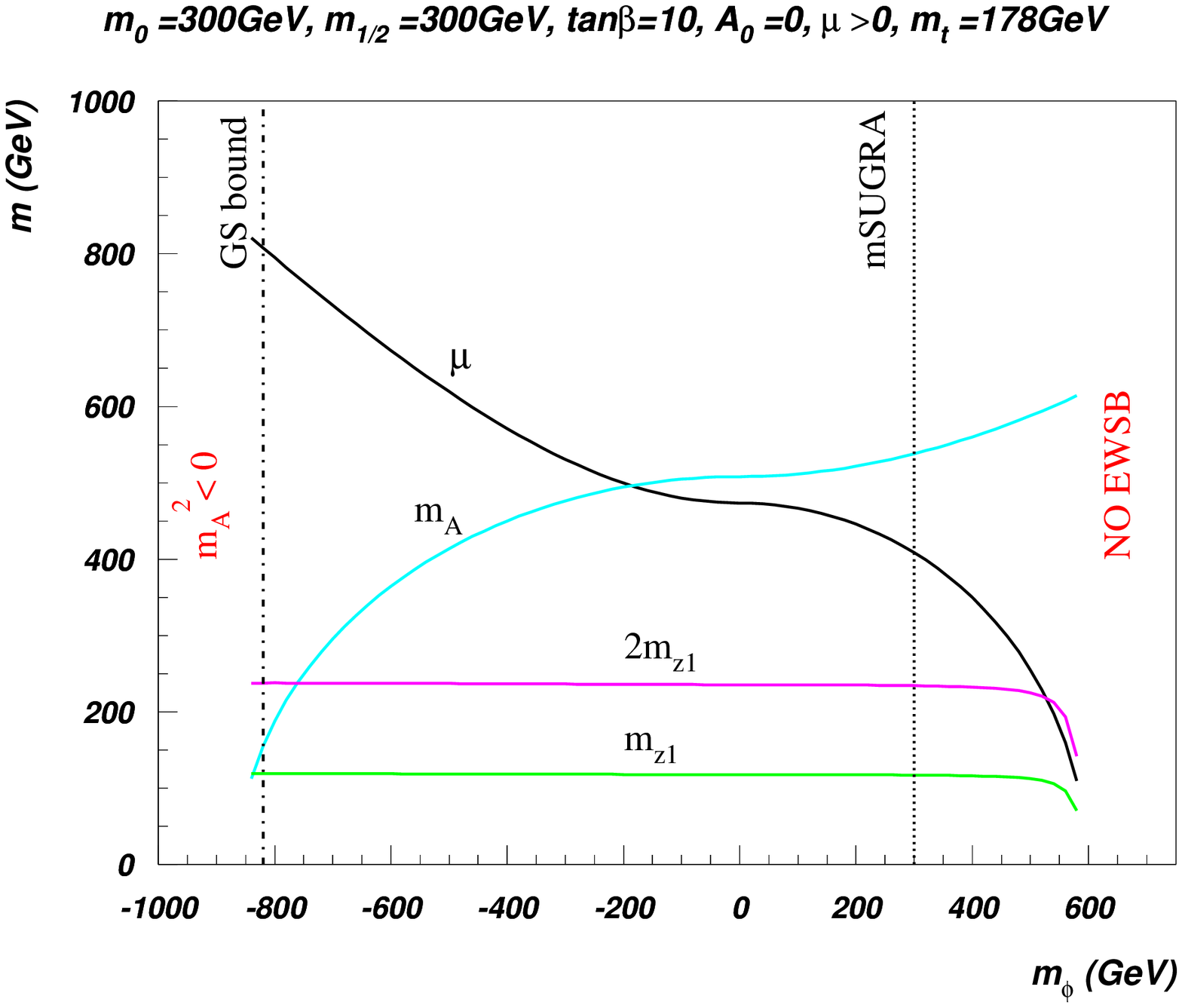,width=12cm} 
\caption{Plot of $\mu$, $m_A$ and $m_{\tz_1}$ vs. $m_\phi$ for
$m_0=300$ GeV, $m_{1/2}=300$ GeV, $A_0=0$, $\tan\beta =10$ and 
$m_t=178$ GeV. We take $\mu >0$.
}
\label{fig:mumavsmH}
}

To understand the behavior of the $\mu$ parameter and $m_A$ in the NUHM1
model, we first examine the renormalization group equations (RGEs) for the soft
SUSY breaking Higgs squared mass parameters. Neglecting the Yukawa
couplings for the first two generations, these read:
\begin{eqnarray}
\frac{dm_{H_u}^2}{dt}&=&\frac{2}{16\pi^2}\left( -{3\over 5}g_1^2M_1^2-
3g_2^2M_2^2+{3\over 10}g_1^2 S+3f_t^2X_t\right),\label{eq:mhu} \\
\frac{dm_{H_d}^2}{dt}&=&\frac{2}{16\pi^2}\left( -{3\over 5}g_1^2M_1^2-
3g_2^2M_2^2-{3\over 10}g_1^2 S+3f_b^2X_b+f_\tau^2X_\tau\right) \label{eq:mhd} ,
\end{eqnarray}
where $t=\log (Q)$, $f_{t,b,\tau}$ are the $t$, $b$ and $\tau$ Yukawa
couplings, and
\begin{eqnarray}
X_t &=& m_{Q_3}^2+m_{\tst_R}^2+m_{H_u}^2+A_t^2 ,\\
X_b &=& m_{Q_3}^2+m_{\tb_R}^2+m_{H_d}^2+A_b^2 ,\\
X_\tau &=& m_{L_3}^2+m_{\ttau_R}^2+m_{H_d}^2+A_\tau^2 ,\ \ {\rm and}\\
S &=& m_{H_u}^2-m_{H_d}^2+ Tr\left[ {\bf m}_Q^2-{\bf m}_L^2 
-2{\bf m}_U^2+{\bf m}_D^2+{\bf m}_E^2 \right] .
\label{eq:S}
\end{eqnarray}
The term $S$ is identically zero in the NUHM1 model, but can be non-zero 
in the NUHM2 model. 

For small-to-moderate values of $\tan\beta$, $f_t\gg f_b,\ f_\tau$, and
so the RGE terms including $X_t$ usually dominate the $X_b$ and
$X_\tau$ terms. The RGE terms including Yukawa couplings occur with
overall positive signs, which results in driving the corresponding soft
Higgs boson mass squared parameters to smaller (and ultimately negative) 
values at the low scale. Indeed, this is
the familiar REWSB mechanism. 
Since $X_t\ni m_{H_u}^2$, a large, positive value of $m_\phi
> m_0$
results in a stronger push of $m_{H_u}^2$ to negative values (relative
to that in mSUGRA) during the
running from $M_{GUT}$ to $M_{weak}$, while the evolution of $m_{H_d}^2$
is rather mild.  Alternatively, if $m_\phi\ll 0$, then there exist
cancellations within the $X_t$ term which results in a milder running of
$m_{H_u}^2$ from $M_{GUT}$ to $M_{weak}$.  Indeed, as shown in
Ref.~\cite{bbmpt}, 
$$\Delta m_{H_{u,d}}^2\equiv
m_{H_{u,d}}^2({\rm NUHM1})-m_{H_{u,d}}^2({\rm mSUGRA})$$ satisfies
\begin{equation}
\Delta m_{H_u}^2({\rm weak})\simeq\Delta m_{H_u}^2({\rm GUT})\times e^{-J_t}\;,
\end{equation}
where $$J_t = \frac{3}{8\pi^2}\int dt f_t^2 > 0,$$ with $f_t$ being the
top quark Yukawa coupling.  We see that $\Delta
m_{H_{u}}$ maintains its sign, but reduces in magnitude 
under evolution from the GUT to the weak scale.
The same argument applies for $\Delta
m_{H_d}^2$, except that the effect of evolution is much smaller because $f_{b,\tau}
\ll f_t$ except when $\tan\beta$ is very large. 
The situation is illustrated
in Fig.~\ref{fig:mHrun}, where we plot the running of $m_{H_u}^2$ and
$m_{H_d}^2$ from $M_{GUT}$ to $M_{weak}$ using the same model parameters
as in Fig.~\ref{fig:mumavsmH}, except for three choices of $m_\phi =500$
GeV, 300 GeV (mSUGRA case) and $-700$ GeV. In these cases, the weak
scale values of $m_{H_u}^2$ are $-(251\ {\rm GeV})^2$, $-(407\ {\rm GeV})^2$ 
and $-(732\ {\rm GeV})^2$, respectively, while the corresponding weak scale
values of $m_{H_d}^2$ are $(527\ {\rm GeV})^2$, $(348\ {\rm GeV})^2$ and
$-(672\ {\rm GeV})^2$.
\FIGURE{\epsfig{file=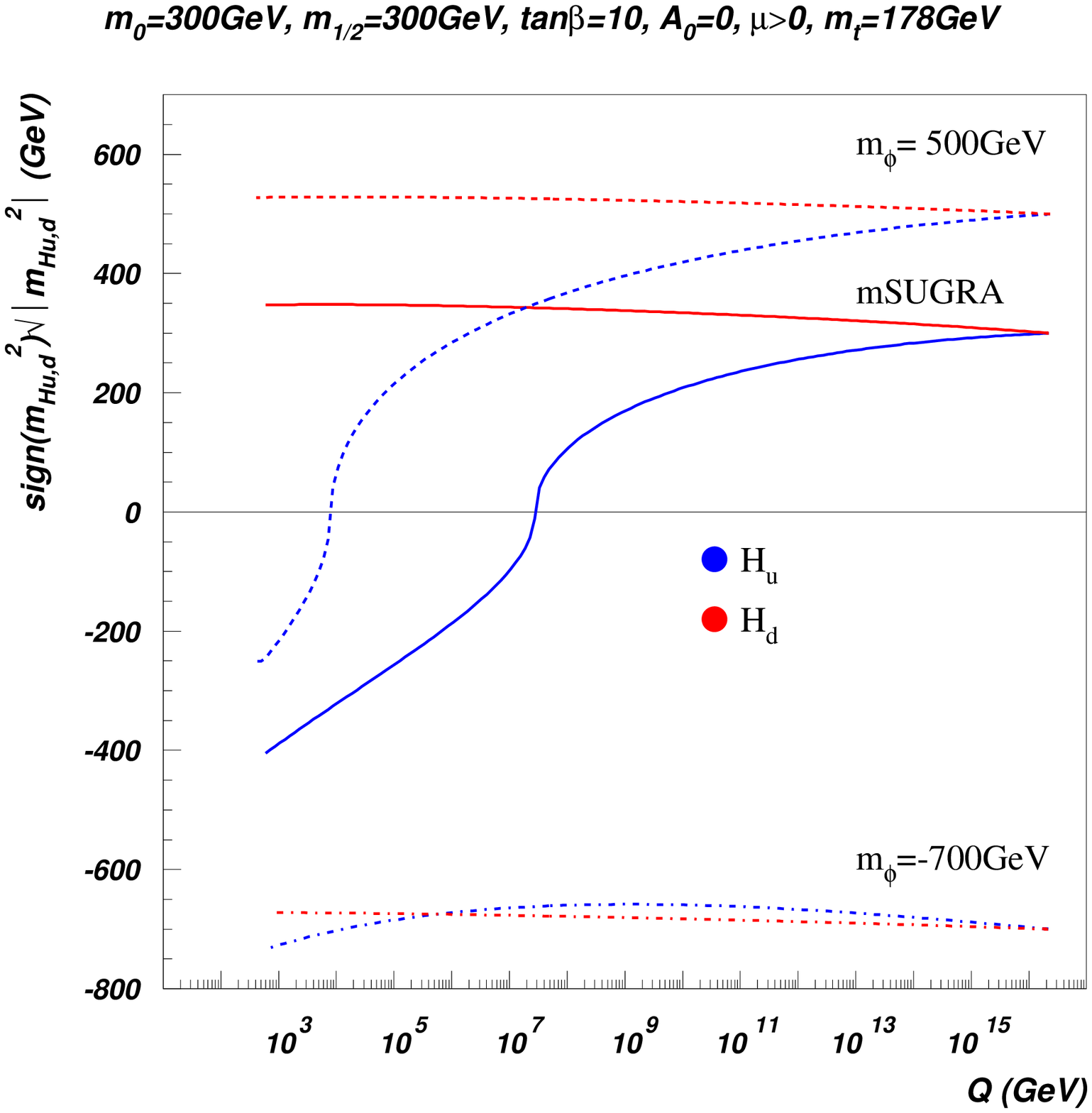,width=12cm} 
\vspace*{-0.8cm}
\caption{The evolution of $m_{H_u}^2$ and $m_{H_d}^2$ from 
$Q=M_{GUT}$ to $M_{weak}$ for $M_\phi =300$ GeV (mSUGRA case), 500 GeV and
-700 GeV. We also take
$m_0=300$ GeV, $m_{1/2}=300$ GeV, $A_0=0$, $\tan\beta =10$ and 
$m_t=178$ GeV, with $\mu >0$.
}
\label{fig:mHrun}
}

The tree level minimization condition for EWSB in the MSSM is
\begin{equation}
\mu^2 =\frac{m_{H_d}^2-m_{H_u}^2\tan^2\beta}{(\tan^2\beta -1)}-
{M_Z^2\over 2} .
\label{eq:ewsb1}
\end{equation}
For moderate to large values of $\tan\beta$ (as favored by LEP2 Higgs boson
mass constraints), and $|m_{H_u}|\gg M_Z$, $\mu^2\sim -m_{H_u}^2$. 
Thus, we see that in the case of large negative $m_\phi$ values, we would
expect a large $|\mu |$ parameter, whereas for large positive $m_\phi$
values, $m_{H_u}^2$ is barely driven to negative values, and we
expect a small $|\mu |$ parameter. Within the same approximation,
the tree level pseudoscalar Higgs mass $m_A$ is given by
\begin{equation}
m_A^2=m_{H_u}^2+m_{H_d}^2+ 2\mu^2\simeq m_{H_d}^2-m_{H_u}^2  .
\label{eq:ewsb2}
\end{equation}
For large negative values of $m_\phi$, the weak scale values of
$m_{H_u}^2$ and $m_{H_d}^2$ are both negative, and can cancel
against the $2\mu^2$ term, yielding small pseudoscalar masses. 
Meanwhile, for large positive values of $m_\phi$, 
$m_{H_d}^2\sim sign(m_\phi ) m_\phi^2$
while $m_{H_u}^2$ is small, but negative. 
In this case 
there is no cancellation
in the computation of $m_A^2$, and  thus we expect
$m_A$ to be large, as shown in Fig.~\ref{fig:mumavsmH}.
\FIGURE{\epsfig{file=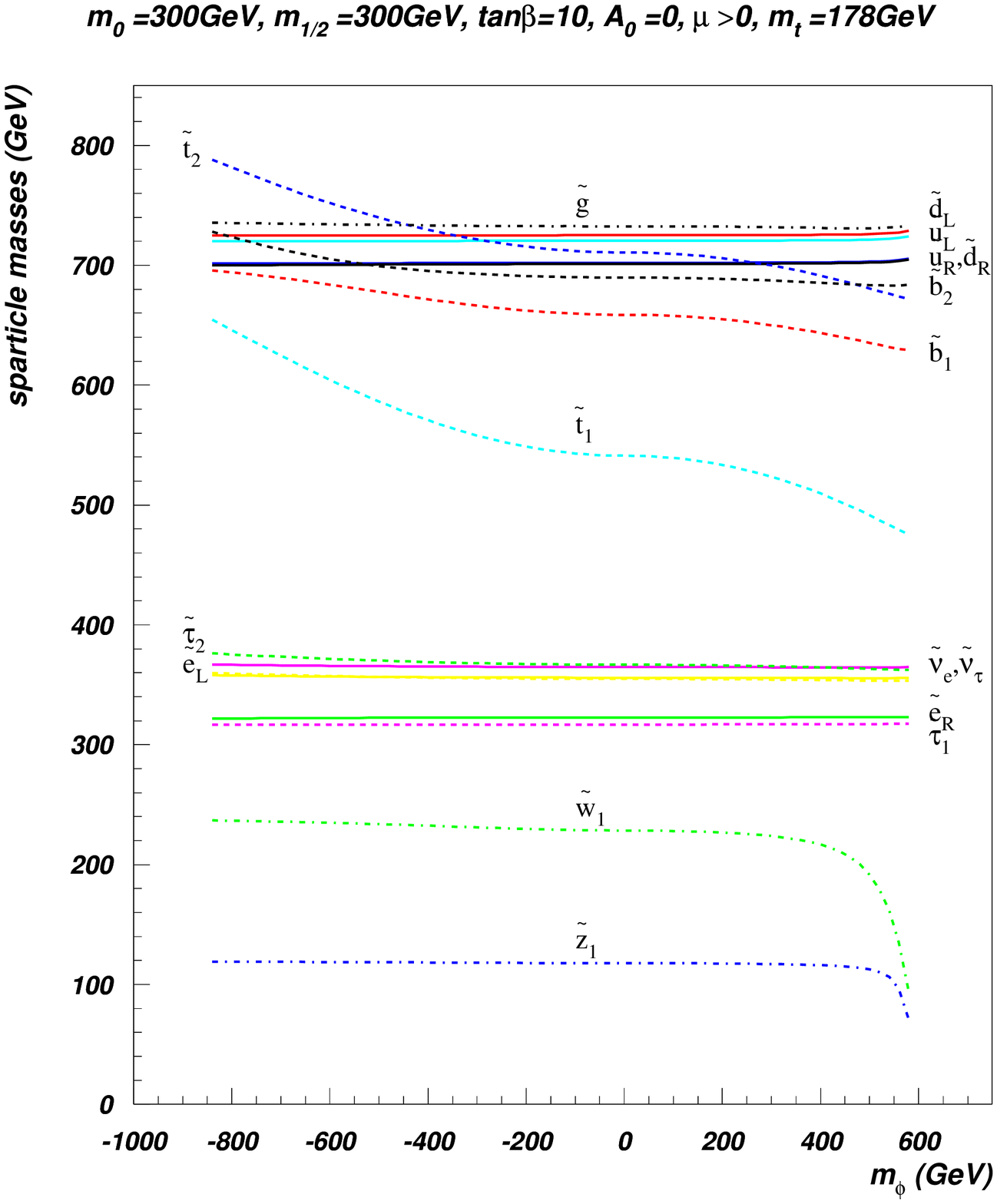,width=12cm,height=10.8cm} 
\vspace*{-0.5cm}
\caption{Various sparticle masses versus $m_\phi$ in the NUHM1 model
for $m_0=300$ GeV, $m_{1/2}=300$ GeV, $A_0=0$, $\tan\beta =10$ and 
$m_t=178$ GeV. We take $\mu >0$.
}
\label{fig:mass_nuhm1}
}

In Fig.~\ref{fig:mass_nuhm1}, we show how different sparticle masses vary
with $m_\phi$ for the same parameter choices as in
Fig.~\ref{fig:mumavsmH}. Most sparticle masses are relatively invariant
to changes in $m_\phi$. One exception occurs for the $\tw_1$ and
$\tz_{1,2}$ masses, which become small when $\mu\alt M_2$, and the
$\tz_1$ becomes increasingly higgsino-like. The other exception occurs
for the $\tst_{1,2}$ and $\tb_{1,2}$ masses. In this case, the
$Q_3\equiv (\tst_L,\tb_L )$ and $\tst_R$ running masses also depend on
terms including $f_t^2 X_t$. Thus, when $X_t$ is small (for $m_\phi <
m_0$), the diagonal entries in the top and bottom squark mass squared
matrices are not as suppressed due to top Yukawa coupling effects as in
the mSUGRA case. In contrast, for large positive values of $m_\phi$,
$X_t$ is large and these soft masses are more suppressed resulting
lighter third generation squarks.  These expectations are indeed born
out in Fig.~\ref{fig:mass_nuhm1}.

\subsection{NUHM1 model: parameter space}
\label{sec:nuhm1_ps}

The mSUGRA parameter space point we have used for illustration so far,
$$(m_0,\ m_{1/2},\ A_0,\ \tan\beta =300\ {\rm GeV}, 300\ {\rm GeV}, 0, 10)$$ 
with $\mu >0$ and $m_t=178$ GeV, is excluded since $\Omega_{\tz_1}h^2 = 1.2
$. However, by extending the parameter space to include $m_\phi$ as in
the NUHM1 model, these parameter values are allowed for an
appropriate choice of $m_\phi$. As an example, in
Fig.~\ref{fig:nuhm1_rscan} we use $m_0, m_{1/2}$ and $A_0$
as in the mSUGRA parameter set above,
and scan over $\tan\beta$ and $m_\phi$ values.  
\FIGURE{\epsfig{file=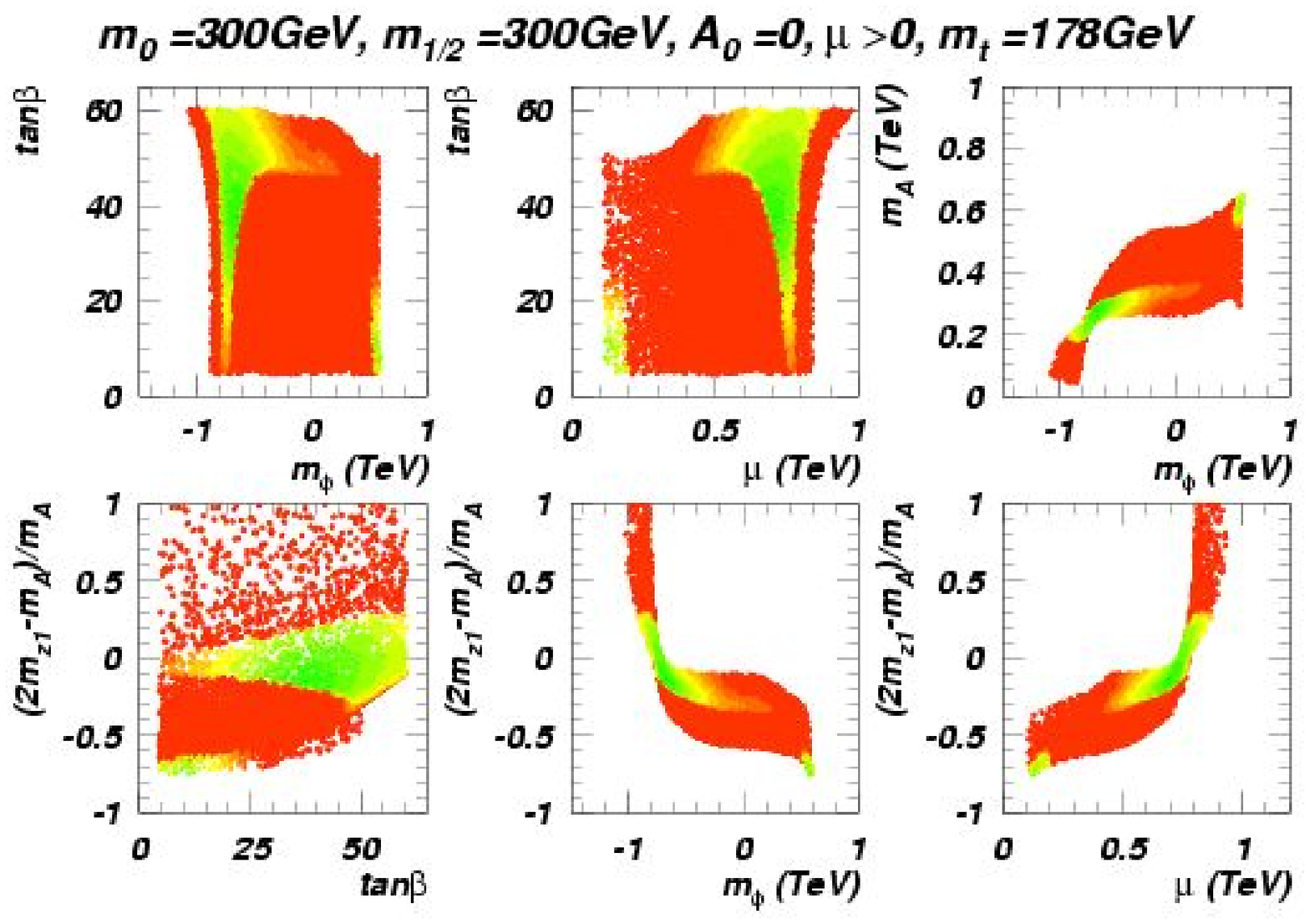,width=14cm} 
\caption{Ranges of $\sqrt{\chi^2}$ in a scan over $\tan\beta$ and
$m_\phi$ values in the NUHM1 model for fixed $m_0=m_{1/2}=300$~GeV,
$A_0=0$, $\mu >0$ and $m_t=178$ GeV. The green region corresponds to low
values of $\sqrt{\chi^2} \alt \sqrt{3}$, while the red region has
$\sqrt{\chi^2}\agt 5$, with the yellow region corresponding to intermediate
values. }
\label{fig:nuhm1_rscan}}
In plotting points, we construct a $\chi^2$ value out of the
three quantities $\Omega_{\tz_1}h^2$, $BF(b\to s\gamma )$ and $\Delta a_\mu$,
where $\chi^2=\sum_{i=1}^{3}\frac{(x_i-\mu_i)^2}{\sigma_i^2}$, where
$x_i$ is the predicted value, $\mu_i$ is the measured value, and
$\sigma_i$ is the error on the $i$th measured quantity.
In constructing the $\chi^2$ value, we only use the WMAP upper
bound (thus, points with $\Omega_{\tz_1}h^2<0.113$ do not contribute to
the $\chi^2$) to allow for the possibility of mixed cold dark matter, 
where for instance a portion of dark matter might consist of, 
say, axions\footnote{A low thermal relic
abundance may also be compatible with a fully supersymmetric dark
matter scenario provided non-thermal production of neutralinos or
cosmological enhancements of the thermal relic density 
occur: see Sec~\ref{sec:dm}.}.
Green
points have low $\sqrt{\chi^2}\alt \sqrt{3}$ values, and agree well with the
central values of each of these measurements. Red points have large
$\sqrt{\chi^2}\agt 5$, and are excluded.  Yellow points have
intermediate values of $\sqrt{\chi^2}$. 
We show six frames illustrating
various correlations amongst parameters. In frame {\it a}) showing
$m_\phi\ vs.\ \tan\beta$, we see the green/yellow $A$-annihilation
funnel for $m_\phi\sim -0.8$ TeV, which occurs for {\it every}
$\tan\beta$ value. 
We also see at $m_\phi\sim 0.6$ TeV the
appearance of the higgsino region, corresponding to the HB/FP region of
the mSUGRA model. While the relic density is in accord with WMAP in this
region, as $\tan\beta$ increases, $BF(b\to s\gamma )$ and $\Delta a_\mu$
also increase, so that the higgsino region becomes increasingly
disfavored for large $\tan\beta$. 
Frame {\it b}) shows the $\tan\beta\ vs.\ \mu$ correlation,
where we see that indeed $\mu$ is small in the higgsino region, and
large in the $A$-funnel. Frame {\it c}) shows the $m_\phi\ vs.\ m_A$
correlation, and indeed we see large values of $m_A$ in the higgsino
region, while $m_A\sim 250$ GeV in the $A$-funnel. The remaining three
frames show $(2m_{\tz_1}-m_A)/m_A\ vs.\tan\beta$, $m_\phi$ and $\mu$
respectively, which explicitly displays the $A$-annihilation funnel
against the input parameters and $\mu$.
\FIGURE{\epsfig{file=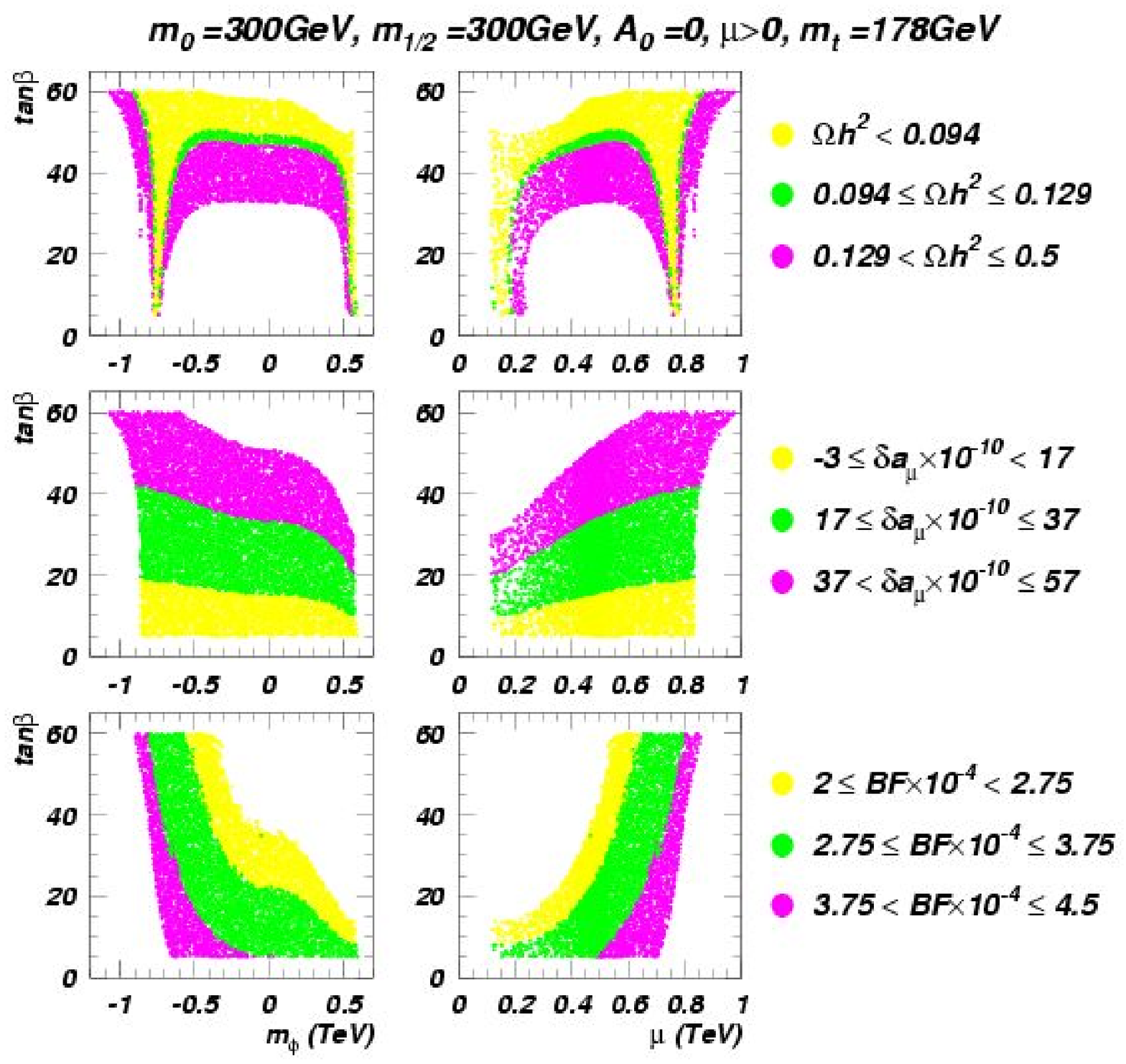,width=13.5cm} 
\vspace*{-0.4cm}
\caption{Ranges of $\Omega h^2$, $\delta a_\mu$ and $BF(b \to s\gamma)$ 
for the NUHM1 models scanned in Fig.~\ref {fig:nuhm1_rscan}.
If a parameter point falls outside the ranges shown, it is not plotted
in this figure.}
\label{fig:nuhm1_rscan2}}

As we have already mentioned, there is still some debate on the range of
the SM prediction for $a_\mu$, as well as the MSSM prediction for $BF(b
\to s\gamma)$.  As a result, the $\chi^2$ values in
Fig.~\ref{fig:nuhm1_rscan} should be interpreted with some care. To
facilitate this, we show the ranges of $\Omega h^2$, $\delta a_\mu$ and
$BF(b\to s\gamma)$ for the same set of NUHM1 models as in
Fig.~\ref{fig:nuhm1_rscan2}. If a parameter set in
Fig.~\ref{fig:nuhm1_rscan} yields a value of these quantities that is
outside the range shown, then the point is not plotted.  From this plot
we see why only the low $\tan\beta$ portion of the higgsino region at
low $\mu$ gives low $\chi^2$ in Fig.~\ref{fig:nuhm1_rscan}: at high
$\tan\beta$, the $BF(b\to s\gamma )$ is quite high, while the value of
$\Delta a_\mu$ is quite low. We see, however, a large swath of yellow
with $\Omega_{CDM}h^2<0.094$ at large values of $\tan\beta$ 
in the upper frames of
Fig.~\ref{fig:nuhm1_rscan2}. This occurs because the $A$ and $H$
bosons become light and relatively wide, 
leading to a resonant enhancement (even off resonance) in the
neutralino annihilation cross section and a corresponding reduction in
the relic density. In this case, there must then be some other 
new physics that brings
the CDM density up to the WMAP value\footnote{We note that it is possible
that this new physics associated with the non-LSP components of dark
matter may also yield (possibly non-calculable) new contributions to
both $\delta a_\mu$ and $BF(b\to s\gamma)$, so that it may be premature
to unequivocally exclude this parameter space region at large values of
$\tan\beta$ because these quantities are not in agreement with
their measured values.}.

\FIGURE{\epsfig{file=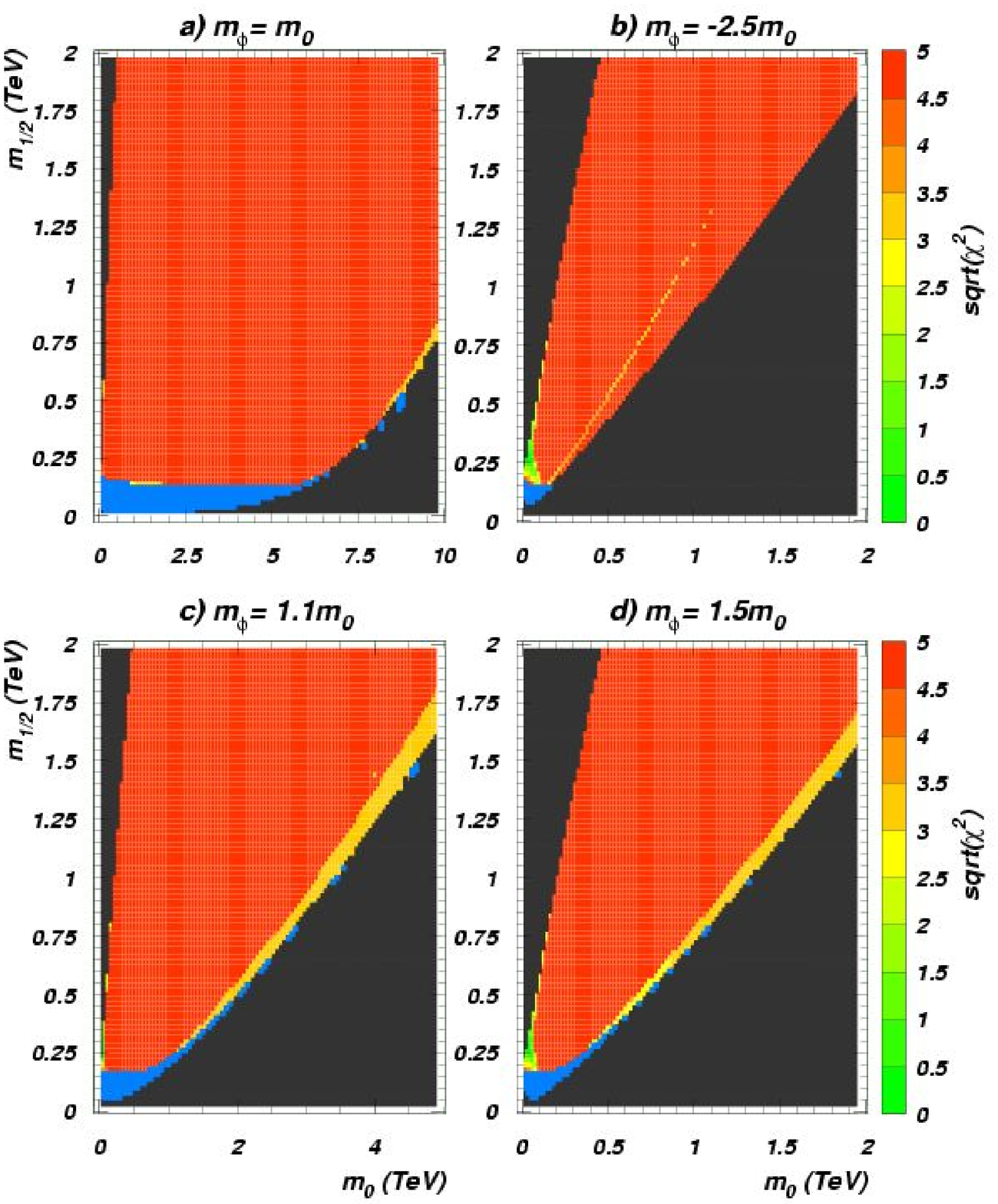,width=13cm,height=13cm} 
\caption{The $\sqrt{\chi^2}$ value in the $m_0\ vs.\ m_{1/2}$ plane for
$A_0=0$, $\tan\beta =10$, $\mu >0$ and $m_t=178$~GeV. In frames {\it
a}), {\it b}), {\it c}), and {\it d}) we take $m_{\phi}/m_0=1,\ -2.5,\
1.1$ and $1.5$, respectively. 
The blue region is excluded by LEP2.
} 
\label{fig:nuhm1_planes}}

In Fig.~\ref{fig:nuhm1_planes}, we show the $m_0\ vs.\ m_{1/2}$
parameter space plane for the NUHM1 model, with $A_0=0$, $\tan\beta
=10$, $\mu >0$, $m_t=178$ GeV and {\it a}) $m_\phi =0$ (mSUGRA case),
{\it b}), $m_\phi =-2.5 m_0$, $m_\phi =1.1 m_0$ and $m_\phi =1.5 m_0$.
The black regions are excluded by lack of REWSB (right hand side) and
because the stau is the LSP on the left hand side.  The blue shaded
region is excluded by the LEP2 constraint that $m_{\tw_1}<103.5$ GeV.
The remaining parameter space is color coded according to the $\sqrt{\chi^2}$
value, and indeed we see that most of parameter space is excluded.  The
mSUGRA case of frame {\it a}) shows the HB/FP region at $m_0\sim 8-10$
TeV, while the stau co-annihilation is squeezed against the left edge of
the allowed parameter space. In frame {\it b}) for a large {\it
negative} value of $m_\phi$, we see that a narrow allowed region now
cuts through the middle of the parameter plane. This is the
$A$-annihilation funnel, which is much narrower than in the mSUGRA case at
large $\tan\beta$, since now the $A$-width is relatively small:
typically $\sim 1$~GeV. Note that the range of $m_0$ extends only to 2
TeV, since for larger values of $m_0$, $m_A^2<0$ and REWSB is violated.
Since $\mu$ remains large, there is no higgsino LSP region along the
right-hand edge of parameter space.  In frame {\it c}) for $m_\phi =1.1
m_0$, we see that the $m_0$ parameter ranges only to 3 TeV, since now
the right-hand side is excluded by $\mu^2<0$.  This leads to a higgsino
LSP region which is shaded yellow, which begins at $m_0\sim 1$ TeV for
low $m_{1/2}$, and explicitly shows that the higgsino LSP region can
occur even for relatively light scalar masses. Frame {\it d}) for
$m_\phi= 1.5 m_0$ shows that the higgsino region has moved to even lower
$m_0$ values, which are below $3$ TeV even for $m_{1/2}$ as high as 2
TeV. 
%

\FIGURE{\epsfig{file=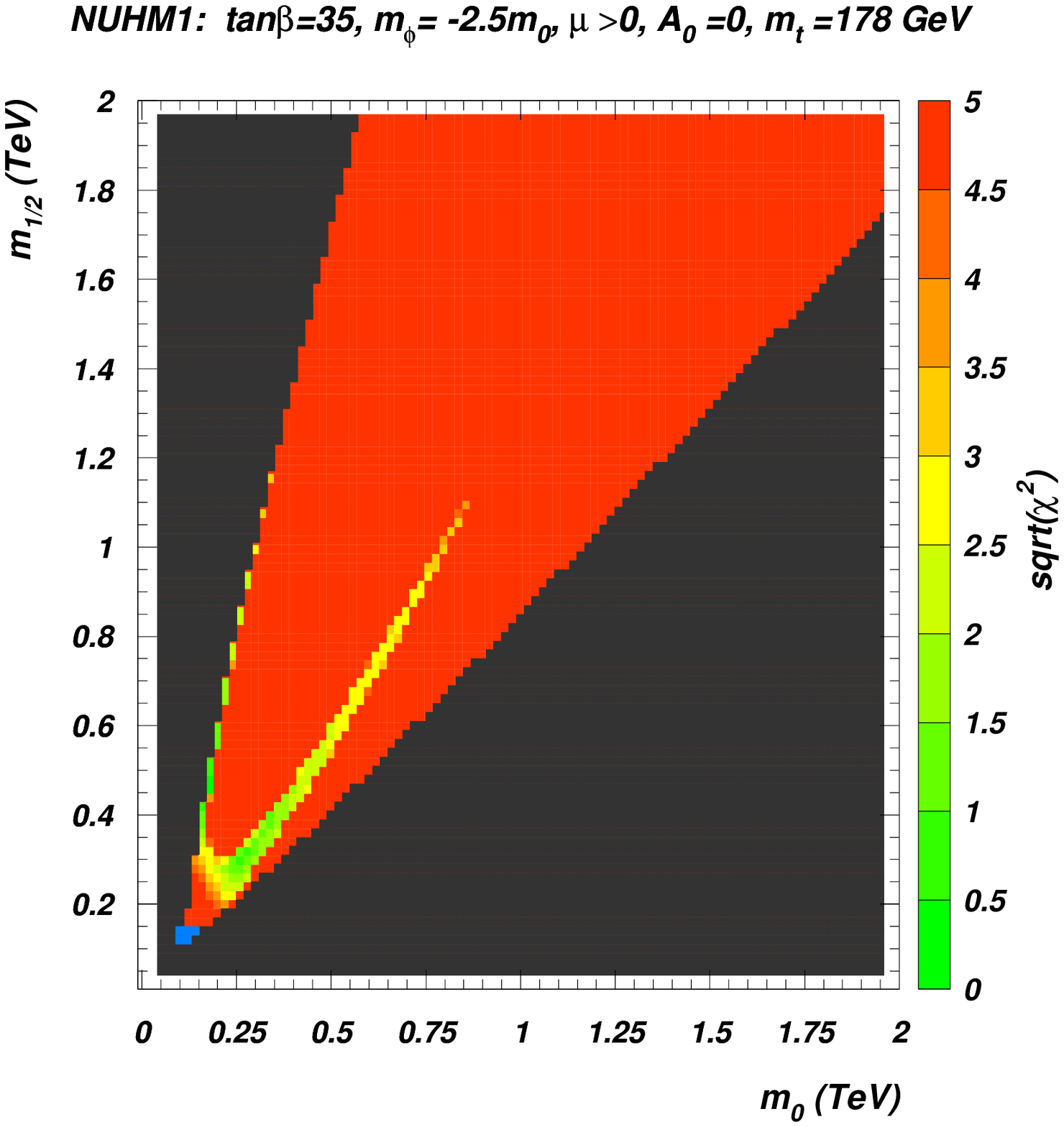,width=13cm,height=13cm} 
\vspace*{-0.5cm}
\caption{The $\sqrt{\chi^2}$ value in the $m_0\ vs.\ m_{1/2}$
plane for $A_0=0$, $\tan\beta =35$, $\mu >0$ and 
$m_t=178$ GeV, with $m_\phi =-2.5 m_0$. 
The blue region is excluded by LEP2.
}
\label{fig:nuhm1_plane35}
}

In Fig.~\ref{fig:nuhm1_plane35}, we show the $\sqrt{\chi^2}$ value in the
$m_0\ vs.\ m_{1/2}$ plane for $A_0=0$, $\tan\beta =35$, $\mu >0$,
$m_\phi =-2.5 m_0$ and $m_t =178$ GeV.  
This plane includes the region
of lowest $\chi^2$ value as indicated in
Fig.~\ref{fig:nuhm1_rscan}. 
Here we see a well-defined $A$-annihilation
funnel for the case of $\tan\beta =35$, where the lower portion gives
excellent agreement to the measured WMAP relic density, the branching
fraction $BF(b\to s\gamma )$ and the muon anomalous magnetic moment
$\Delta a_\mu$. Excellent agreement is also obtained in this case for the
stau co-annihilation region for $m_{1/2}\sim 350-600$ GeV.

\subsection{Dark matter detection: overview and methodology}
\label{sec:dm}

Two major issues enter the evaluation of the prospects for 
neutralino dark matter detection and of assessing the relative 
effectiveness of the various direct and indirect techniques that have
been proposed:

\begin{enumerate} 
\item Since the detection rates critically depend on the dark matter 
halo profile of our own galaxy, and a wealth of observational data 
and theoretical constraints are available, the halo models one resorts 
to must be {\em consistent} with all information we have about the 
Milky Way.  Moreover, since different detection techniques rely 
on the local dark matter density distribution and on the 
velocity distribution, a fair comparison among them may be 
carried out only provided the two distributions are {\em self-consistently} 
computed, not only locally, 
but throughout the whole halo\footnote{We recall that, while the 
gamma ray flux from the galactic center only depends on the 
dark matter density in a very small, and poorly known region, 
the antimatter rates vary with the dark matter density in a more 
extended portion of the halo; finally the flux of neutrinos from 
neutralino annihilations in the sun and the direct detection rates 
essentially depend only on the density and velocity distributions 
of the dark halo within the solar system.}.

\item A comparison of the various techniques must rely on quantities
which provide information on the relative strength of the expected
signal with respect to the projected experimental sensitivity, taking
into account the background.  Toward this end, we use what have been
dubbed {\em Visibility Ratios} (VR)
~\cite{Profumo:2004ty,Profumo:2004at,Masiero:2004ft,bbmpt}, which
will be defined below for each experiment.  A VR is simply a {\em
signal-to-sensitivity} ratio: when VR$>$1, the signal calculated using a 
specified model is expected to be detectable over backgrounds
with the particular
experimental setup; in case VR$<$1, the signal will lie below the
sensitivity of the considered detection facility. The locus of points at
VR=1 outlines the actual {\em reach contour} of any particular detection
technique. The relative magnitudes of the VR's for different dark matter
search experiments provides a direct comparison 
between them\cite{profumo_in_prep}.
\end{enumerate}

As far as item 1. is concerned, we follow here the strategy outlined 
in Ref.~\cite{Profumo:2004ty} (the reader is referred to 
Ref.~\cite{pierohalos,Edsjo:2004pf,Profumo:2004at} for more details). 
Motivated by N-body simulations of hierarchical clustering in dark matter 
cosmologies, the starting point is to take into account the correlation 
between the virial mass of a given galaxy $M_{vir}$ and its concentration 
parameter $c_{vir}$~\cite{NFW}, and then to provide a description of the 
dynamics of the baryon infall and of its back-reaction onto the dark halo. 
Two extreme regimes have been outlined, the first one being motivated by 
a large angular momentum transfer between dark matter and baryons, 
while the second one by supposing that baryons settled in with no net 
transfer of angular momentum to the dark component. 

In the first scenario, the central cusp in the dark matter halo, 
as seen in numerical simulations, is smoothed out by a significant 
heating of the cold particles~\cite{elzant}, leading to a cored 
density distribution, which has been modeled by the so called 
{\em Burkert profile}~\cite{burkert},
\begin{equation}\label{eq:burkert}
\rho_B(r)=\frac{\rho_B^0}{(1+r/a)\left(1+(r/a)^2\right)}.
\end{equation}
Here, the length scale parameter 
has been set to $a=11.7$ kpc, while the normalization
$\rho_B^0$ is adjusted to reproduce the local halo density at the 
Earth position to $\rho_B(r_0)=0.34 \ {\rm GeV\ cm}^{-3}$~\cite{pierohalos}. 
We refer to this model as to the {\em Burkert Halo Model}.
It has been successfully tested against a large sample of rotation 
curves of spiral galaxies~\cite{salucci}. 

In the second scenario that we consider, 
baryon infall causes a progressive 
deepening of the gravitational potential well at the center of the galaxy, 
resulting in increasingly higher concentration of dark matter particles. 
In the circular orbit approximation~\cite{blumental,ulliobh}, 
this adiabatic contraction limit has been worked out starting from the 
N03 profile proposed in Ref.~\cite{n03}; the resulting spherical profile, 
which has no closed analytical form, roughly follows, in the inner 
galactic regions, the behavior of the profile proposed by 
Moore {\it et al.},\cite{moore}, approximately scaling as $r^{-1.5}$ in the 
innermost regions, and features a local dark matter density 
$\rho_{N03}(r_0)=0.38 \ {\rm GeV\ cm}^{-3}$. 
We dub this setup as the {\em Adiabatically Contracted N03 Halo Model}.

The parameters for both models have been chosen to reproduce a variety 
of dynamical information, ranging from the constraints 
stemming from the motion of stars in the sun's neighborhood, 
total mass estimates from the motion of the outer satellites, 
and consistency with the Milky Way rotation curve and measures of the 
optical depth toward the galactic bulge~\cite{pierohalos,Edsjo:2004pf}.
Finally, the local velocity distributions of the two spherical 
profiles have been self-consistently computed using the formalism 
outlined in Ref.~\cite{eddington}, thus allowing for a reliable comparison 
between direct and indirect detection techniques. 
Both models have been included 
in the latest public release of the {\tt DarkSUSY} 
package~\cite{Gondolo:2004sc}.

Turning to item 2., we begin by defining the 
Visibility Ratios (VR) mentioned above for each dark matter detection 
technique.
For direct dark matter detection, in view of the fact that 
spin-dependent searches have been shown to be sensitive to 
scattering cross sections typically much larger than those predicted 
within the MSSM~\cite{Profumo:2004at}, we focus here only 
on spin-independent (SI) searches. The relevant quantity is the 
neutralino-proton SI scattering cross section, 
$\sigma_{\widetilde Z_1 p}^{\rm SI}$.
The VR is simply defined as the ratio between the expected
$\sigma_{\widetilde Z_1 p}^{\rm SI}$ from a given supersymmetric model
and the corresponding experimental sensitivity $\sigma^{\rm SI}_{\rm
exp}$ at the given neutralino mass, for the particular dark matter halo
profile under consideration, taking into account the corresponding
self-consistently computed local velocity distribution.\footnote{The
sensitivity of direct detection experiments may vary over more than two
orders of magnitude for the typical range of neutralino masses in the
MSSM. It also depends (less strongly) on the halo profile. In view of
this sensitivity, we must be careful to use the corresponding value of
$\sigma^{\rm SI}_{\rm exp}(m_{\widetilde Z_1})$ in the computation of
the VR.}  
We consider here two benchmark direct detection experiments,
namely CDMS-II~\cite{Akerib:2004fq} (stage-2 detectors) and the proposed
XENON 1-ton facility (stage-3 detectors)~\cite{Aprile:2004ey,hbdm}.

To assess the sensitivity of a ${\rm km}^2$-size detector 
such as IceCube designed to detect
high energy muons, we compute the expected muon flux 
from neutralino annihilations in the 
sun\footnote{We also evaluated the muon flux from the Earth, 
but always found far smaller fluxes than those expected from the sun.} 
above a threshold of 1 GeV.
The low threshold was chosen to allow a comparison of the expected muon
flux with the existing upper limits from the SuperKamiokande experiment
(which indeed has a much lower energy threshold than IceCube) and to ensure
that the model is not already excluded. 
The predicted muon flux is then compared with the sensitivity of IceCube.
Since this detector has a higher energy threshold $\sim 50$~GeV, we
corrected for the threshold mismatch using the corresponding projected
sensitivities worked out in Ref. \cite{icecube}. 
We have checked that the projected 
maximal sensitivity to the muon flux obtained using this procedure is in 
good agreement with that obtained by integrating above the actual IceCube 
threshold. 
We also incorporated the dependence of the sensitivity on the soft ({\em e.g.}
from neutralino annihilation to $b\bar b$) and hard ({\em e.g.} 
from annihilation to $W^+W^-$, $Z^0Z^0$ and $t\bar t$) neutrino
spectra--- the sensitivity being smaller in the case of the softer spectrum.

Turning to antimatter experiments, we compute the solar-modulated 
positron, antiproton and antideuteron fluxes, following the procedure 
outlined in Ref.~\cite{Profumo:2004ty}. 
We calculate the neutralino annihilation rates to $\bar{p}$ and
$\bar{n}$ 
using the 
{\tt Pythia} 6.154 Monte Carlo code~\cite{pythia} as implemented in 
{\tt DarkSUSY}~\cite{Gondolo:2004sc}, and then deduce the $\overline{D}$ 
yield using the prescription suggested in Ref.~\cite{dbar}.
The propagation of charged 
cosmic rays through the galactic magnetic fields is worked out through 
an effective two-dimensional diffusion model in the steady state 
approximation, while solar modulation effects were implemented through the 
analytical force-field approximation of 
Gleeson and Axford~\cite{GleesonAxford}.
The solar modulation parameter $\Phi_F$ is computed from the proton 
cosmic-ray fluxes, and assumed to be charge-independent. 
The values of $\Phi_F$ we use for antiprotons and positrons refer to a 
putative average of the solar activity over the three years of data 
taking of the PAMELA experiment~\cite{Morselli:2003xw}, which will be the 
first space-based experiment for antimatter searches. 

For antideuterons
we consider the reach of the proposed gaseous antiparticle 
spectrometer (GAPS)~\cite{Mori:2001dv} proposed to be 
placed on a satellite orbiting around the Earth, and tuned to look for 
antideuterons in the very low kinetic energy interval from 0.1 
to 0.4 GeV~\cite{Mori:2001dv,Hailey:2003xh}. In this energy interval, 
the estimated background is strongly suppressed~\cite{dbar,donato-pbar}, 
and unambiguous evidence of even a single low energy antideuteron could be 
regarded as a positive search result. Since the experiment is expected 
to be launched close to the maximum of the solar cycle,
we set the value of the solar modulation 
parameter $\Phi_F$ at the corresponding value.  
The resulting sensitivity of 
GAPS has been determined to be of the level of 
$2.6\times 10^{-9} {\rm m}^{-2}{\rm sr}^{-1} {\rm GeV}^{-1} {\rm s}^{-1}$
~\cite{Mori:2001dv}. The VR for antideuterons will then be given by the 
integrated $\overline{D}$ flux over the kinetic energy interval 
$0.1<T_{\overline{D}}<0.4$ GeV, divided by the GAPS sensitivity. 
In other words, the VR will give, in this case, the actual number of 
antideuterons expected to be detected by GAPS. 

Regarding antiproton and positron fluxes, it was realized long ago
that the low energy tails are considerably modified by solar modulation
effects and, in the case of antiprotons, also by large secondary and
tertiary backgrounds~\cite{Bergstrom:1999jc}, so that low energy
positrons and antiprotons cannot provide a clean test for new physics
contributions. However, in view of the fact that novel space-based
experiments will be able to extend the experimental sensitivity to these
to the hundreds of GeV range,  the best place
to look for an antimatter signal from neutralino annihilations lies at
the high energy end. Moreover, a distinctive signature is
provided by the clean cutoff of the neutralino-induced
antimatter flux corresponding to energies equal to the neutralino mass,
provided this flux is large enough to be detectable.  Nevertheless, in
general, the
energy spectra of antiparticles generally look quite featureless: aside 
from the fact that the antiparticles from neutralino annihilation 
sit on top of a
featureless background, solar modulation and propagation effects also
tend to wash away any features inherent to the spectrum.
With some obvious exceptions, the antiparticle energy (or rather
$E/m_{\tz_1}$) where the signal-to-background ratio is largest, is
sensitive to the composition of the neutralino
~\cite{Baltz:1998xv,profumo_in_prep}.
Restricting attention to particular energy bin(s) may also be misleading
because of large uncertainties from secondary and tertiary $\bar{p}$.
A possible exception may be that of neutralinos mainly annihilating into
gauge bosons, in which case one expects a bump in the positron spectrum
at approximately half the neutralino mass~\cite{Feng:2000zu}; this could
be especially relevant for neutralinos in the HB/FP region.

To circumvent these problems, we adopt the statistical treatment of the
antimatter yields introduced in Ref.~\cite{Profumo:2004ty} (an analogous
approach has been proposed for cosmic positron searches
~\cite{Hooper:2004bq}). Motivated by the fact that the signal is much
smaller than the background, we introduce a quantity which weighs its
``statistical significance, summed over the energy bins'',
\begin{equation}\label{eq:iphi}
I_\phi=\int_{T_{\rm min}}^{T_{\rm max}}\frac{\left(\phi_s(E)\right)^2}
{\phi_b(E)}{\rm d}E,
\end{equation}
where $\phi_s(E)$ and $\phi_b(E)$ respectively represent the antimatter 
differential fluxes 
from neutralino annihilations and from the background at antiparticles' 
kinetic energy $E$, and $T_{\rm min,\ max}$ 
correspond to the antiparticle's maximal and minimal kinetic energies to which 
a given experiment is sensitive (in the case of the PAMELA 
experiment~\cite{Morselli:2003xw}, 
$T^{e^+}_{\rm min}=50$ MeV, $T^{e^+}_{\rm max}=270$ GeV, 
$T^{\bar p}_{\rm min}=80$
MeV and $T^{\bar p}_{\rm max}=190$ GeV). 
We compute the primary component, $\phi_s$, with the {\tt DarkSUSY} package, 
interfaced with a subroutine implementing the diffusion and solar modulation 
models outlined above. The background flux $\phi_b$ has been calculated with 
the {\tt Galprop} package~\cite{galprop}, with the same parameter choices 
employed to compute the signal. 

Given an experimental facility with a geometrical factor
(acceptance) $A$ and a total data-taking time $T$, it has been 
shown~\cite{Profumo:2004ty} that, in the limit of a large number of energy 
bins and of high precision secondary ({\em i.e.} background) flux 
determination, a SUSY model giving a primary antimatter flux $\phi_s$ 
can be discriminated at the 95\% C.L. if
\begin{equation}
I_\phi(\phi_s)\cdot A\cdot T>(\chi^2)^{95\%}_{\rm n_b},
\end{equation}
where $(\chi^2)^{95\%}_{\rm n_b}$ stands for the 95\% C.L. $\chi^2$ with 
${\rm n_b}$ degrees of freedom. 
For the PAMELA experiment, where $A=24.5\ {\rm cm}^2\ 
{\rm sr}$, $T$=3 years and ${\rm n_b}\simeq60$ we get the following 
discrimination condition~\cite{Profumo:2004ty}
\begin{equation}
I_\phi(\phi_s)>\frac{(\chi^2)^{95\%}_{\rm n_b}}{A\cdot T}
\equiv I_\phi^{\rm 3y,\ PAMELA,\ 95\%}\simeq3.2\times 10^{-8} 
\ {\rm cm}^{-2} {\rm sr}^{-1} {\rm s}^{-1}
\end{equation}
which is approximately valid for both positrons and antiprotons 
(though in the latter case the PAMELA experiment 
is expected to do slightly better). 
As a rule of thumb, the analogous quantity for AMS-02 should improve at 
least by one order of magnitude~\cite{Feng:2000zu}. 
We therefore define as VR for antiprotons and for positrons the ratio
\begin{equation}
({\rm VR})^{\bar p,e^+}\equiv I_\phi^{\bar p,e^+}/
I_\phi^{\rm 3y,\ PAMELA,\ 95\%}.
\end{equation}


Finally, the case for the gamma ray flux from the galactic
center is plagued by large uncertainties on the very central structure
of the Milky Way dark halo.  Depending on
various assumptions on the galactic models and on the physical cut-off
in the inner part of our galaxy, there might be a spread of various
orders of magnitude in the computation of the actual 
gamma-ray flux\cite{Fornengo:2004kj,bo}.  This may be written as a product
of a purely astrophysical quantity describing the propagation of the
photons to the detector, and 
of a purely particle-physics
quantity $\langle \sigma v\rangle$~\cite{Bergstrom:1997fj} describing
the source of these photons.  The former
reads,
\begin{equation}
\langle J(0)\rangle(\Delta\Omega)=\frac{1}{\Delta\Omega}
\int_{\Delta\Omega}J(\psi){\rm d}\Omega,
\end{equation}
where
\begin{equation}
J(\psi)=\frac{1}{8.5\ {\rm kpc}}\cdot\left(\frac{1}{0.3\ 
{\rm GeV}/{\rm cm}^3}\right)^2
\int_{\rm line\ of\ sight}\rho^2(l){\rm d}l(\psi).
\end{equation}

The attitude we take here is just to extrapolate the halo models we use
in the $\rho(r\rightarrow0)$ limit, and to compute the corresponding
$\langle J(0)\rangle$ for the acceptance $\Delta\Omega$ of GLAST. What
we find is $\langle J(0)\rangle=7.85$ for the Burkert Halo Model, and
$\langle J(0)\rangle=1.55\times10^{5}$ for the Adiabatically Contracted
N03 Halo Model.\footnote{In the case of the Adiabatically Contracted halo
model, this procedure is quite arbitrary, since different hypotheses on
the dynamics of the central black hole formation might lead to very
different predictions for the dark matter density in the center of the
halo~\cite{ulliobh,milo}. Because there is an unconstrained
extrapolation, essentially any flux may be possible.}  
We then compute the integrated gamma ray flux
above a 1 GeV threshold, $\phi^\gamma$, and define the corresponding VR
as $\phi^\gamma/(1.5 \times 10^{-10}\ {\rm cm}^{-2}{\rm s}^{-1})$, the
latter being the corresponding estimated sensitivity of the GLAST
satellite~\cite{Cesarini:2003nr}.

\subsection{Dark matter detection: the NUHM1 model}\label{sec:dm1}

In Fig.~\ref{fig:dm_NUHM1_1} and \ref{fig:dm_NUHM1_2} we show the VR's
for the various experiments that we detailed in Sec.~\ref{sec:dm}, for
two representative mSUGRA parameter choices: we fix $\tan\beta=10$,
$m_{0}=300$ GeV, $m_{1/2}=300$ GeV in Fig.~\ref{fig:dm_NUHM1_1}, and
$\tan\beta=20$, $m_{0}=1000$ GeV, $m_{1/2}=200$ GeV in
Fig.~\ref{fig:dm_NUHM1_2} (in both cases $A_0=0$, ${\rm sign}(\mu )>0$ and
$m_t =178$ GeV) and show the results as a function of
$m_\phi/m_0$.  The regions shaded in red are excluded by the LEP2 limits
on the mass of the pseudoscalar Higgs boson $m_A$ and on the mass of the
lightest chargino. The green regions indicate, instead, parameter space
regions where the neutralino relic abundance $\Omega_{\widetilde Z_1}
h^2<0.13$, consistently with the WMAP 95\% C.L. upper limit on the Cold
Dark Matter abundance~\cite{Spergel:2003cb}: agreement with the central
value of WMAP is obtained close to the boundary of this region.  We
remind the reader that a VR larger than unity means that the signal
should be detectable in the particular dark matter detection channel.
For definiteness, we adopt the conservative Burkert Halo Model.
The results we show should be regarded as plausible {\em
lower limits}, particularly as far as indirect rates are concerned
since a cuspy inner dark halo would greatly enhance the dark matter
detection rates~\cite{Profumo:2004ty,Fornengo:2004kj}.\footnote{This is
  explicitly illustrated for the NUHM2 Model in the next Section.}

We see, in Fig.~\ref{fig:dm_NUHM1_1}, that for all values of $m_\phi$
the signal will be accessible to stage-3 direct detection facilities,
like XENON 1-ton~\cite{Aprile:2004ey}, while stage-2 detectors (such as
CDMS-II\cite{Akerib:2004fq}) will be able to probe only the HB/FP
region, at large $m_\phi$. The behavior of the direct detection VR's is
readily understood. On the one hand, when $m_\phi$ takes large negative
values, both $m_A$ and the $CP$-even heavy Higgs boson mass $m_H$
decrease; this leads to an enhancement of the $t$-channel $H$ exchange
in the neutralino-proton cross section (which scales as $m_H^{-4}$,
assuming that $m_{A,H} \gg m_{\tz_1}$).  On the other hand, when
$m_\phi$ is large and positive, the higgsino fraction increases, and so
does $\sigma_{\tz_1 p}^{\rm SI}$ since 
$\sigma_{\tz_1 p}^{\rm SI}\propto (Z_h Z_g)^2$, where $Z_{h,g}$
respectively denote the higgsino and bino fraction in the lightest
neutralino.  The same behavior for the direct detection VR applies to
Fig.~\ref{fig:dm_NUHM1_2}; here, once again, stage-3 detectors will be
able to fully explore the parameter space, while only the focus point
region will be discoverable at stage-2 facilities.

Turning to the neutralino-annihilations-induced flux of muons from the
sun, we see that in both figures the resonant annihilation region
gives rates which will be various orders of magnitude {\em below} the
projected ultimate sensitivity of IceCube. This is because the
neutralino capture rate in the sun (which mainly depends on the
neutralino-proton  {\em spin-dependent} scattering cross section) is not
large enough:
in contrast to the spin-independent cross section, in fact, the main
contribution to the spin-dependent one comes from the $Z$ exchange
diagram, which is not enhanced by the smaller values of $m_{H,A}$. On
the other hand, a larger higgsino fraction and annihilation cross
section yields detectable rates at neutrino telescopes for the model
considered in Fig.~\ref{fig:dm_NUHM1_1}, at large $m_\phi$; for the case
addressed in Fig.~\ref{fig:dm_NUHM1_2} we see that the expected rates
only lie less than one order of magnitude below the maximal sensitivity
of IceCube. This is because for the smaller value of $m_{1/2}$, the
neutralino LSP does not acquire a sufficiently large higgsino component
all the way to the LEP2 limit. 

As far as other indirect detection techniques are concerned, though we
are here considering the conservative cored dark matter profile, the
large neutralino pair annihilation rate $\langle \sigma v\rangle_0$ in
the funnel region yields very large rates in all channels, peaked on the
value of $m_\phi$ at which $m_A\simeq 2 m_{\widetilde Z_1}$.  A
considerable enhancement is also seen in the HB/FP region.  Despite this
enhancement, antimatter detection rates might not be large enough to
be discriminated against the background, especially in the case that we study
in Fig.~\ref{fig:dm_NUHM1_2}.

\FIGURE{\epsfig{file=dm_NUHM1_1.eps,width=16cm}
\caption{Dark matter detection Visibility Ratios ({\em i.e.}
signal-to-projected-sensitivity ratios) for various direct and indirect
techniques, as a function of the GUT scale non-universal Higgs mass
parameter $m_\phi$, for mSUGRA parameters $\tan\beta=10$, $A_0=0$, ${\rm
sgn}\mu>0$, $m_{0}=300$ GeV, $m_{1/2}=300$ GeV, and setting $m_t=178$
GeV.  We take the dark matter distribution to be given by the cored
Burkert Halo profile described in the text.  A Visibility Ratio larger
than one means that the signal will be detectable over background.
}
\label{fig:dm_NUHM1_1}
}

We should mention that for ranges of parameters that yield a thermal
dark matter density smaller than the WMAP central value we do not
correspondingly scale down the results in Fig.~\ref{fig:dm_NUHM1_1}
and Fig.~\ref{fig:dm_NUHM1_2}, as we would have to for any model where
the remainder of the dark matter was composed of something other than
the SUSY LSP (see e.g. Ref.~\cite{Bottino:2001it} and
references therein). In this case, possible additional contributions
from these other dark matter components would have to be included.
We work here, instead, under the hypothesis that even within the
low thermal relic density parameter ranges, the LSP is all of the
dark matter, but that there is either additional
{\em non-thermal production}
~\cite{non-therm}, or  {\em cosmological relic density enhancement},
as envisaged in Ref.~\cite{quint,Profumo:2003hq} for
quintessential cosmologies, in Ref.~\cite{Catena:2004ba} for
Brans-Dicke-Jordan cosmologies, and in Ref.~\cite{shear,Profumo:2004ex}
for anisotropic cosmologies.
%
\FIGURE{
\epsfig{file=dm_NUHM1_2.eps,width=16cm}
\caption{The same as Fig.~\ref{fig:dm_NUHM1_1}, 
but with mSUGRA parameters $\tan\beta=20$, $A_0=0$, 
$\mu>0$, $m_0=1000$ GeV, $m_{1/2}=200$ GeV.}
\label{fig:dm_NUHM1_2}
}

\subsection{NUHM1 model: Collider searches for SUSY}
\label{sec:nuhm1_col}

While direct and indirect detection techniques that we have just
discussed could establish
the existence of dark matter, 
collider experiments would be needed to make the link with
supersymmetry~\cite{colcos}. 
Collider expectations within the NUHM1 framework can differ from
corresponding expectations within the well-studied mSUGRA model.
In the following discussion, we highlight these differences
confining our discussion to NUHM1 model
parameter sets that satisfy the WMAP bound on $\Omega_{\tz_1}h^2$.

\subsubsection{Fermilab Tevatron}

The most promising avenue for a supersymmetry discovery at the 
Fermilab Tevatron in the case of gravity-mediated SUSY breaking models
with gaugino mass unification is by the observation of trilepton signals from 
$p\bar{p}\to\tw_1\tz_2 X$ followed by $\tw_1\to \ell\nu_\ell\tz_1$ and
$\tz_2\to \ell\bar{\ell}\tz_1$ three body decays, where $\ell= e$ 
or $\mu$~\cite{trilep,baertev}.
In the case of the NUHM1 model where $m_\phi$ is taken to have negative
values so that neutralinos annihilate via the $A$ and $H$ poles, 
the only effect on the gaugino/higgsino sector is that the 
magnitude of the $\mu$ parameter increases. As a result,
the HB/FP region of small $|\mu|$ is absent in frame {\it b}) of
Fig.~\ref{fig:nuhm1_col}, so that unlike the case of mSUGRA~\cite{baertev},
probing large values of $m_0$ and
$m_{1/2}$ via this channel will not be possible. 
Alternatively, if $m_\phi$ is taken to be large compared to $m_0$
so that $|\mu |$ is
small, then chargino and neutralino masses can become lighter, which may
increase production cross sections. Furthermore, the
$m_{\tz_2}-m_{\tz_1}$ mass gap will diminish, which can close
``spoiler'' decay modes such as $\tz_2\to\tz_1 h$, so that the necessary
three-body neutralino decays are more likely to be in effect, and
because $\tan\beta$ is not necessarily large, we do not expect events
with tau leptons to dominate at the expense of $e$ and $\mu$ events.
Thus, in
the large $m_\phi$ region, we expect improved prospects for clean
trilepton signals. Detailed simulation would of course be necessary to
draw definitive conclusions.

\subsubsection{CERN LHC}

The CERN LHC is expected to begin operation in 2007 with
$pp$ collisions at $\sqrt{s}=14$ TeV.
In most regions of parameter space of gravity-mediated SUSY breaking models, 
gluino and squark pair production is expected to be the dominant
source of sparticles at the LHC. Since the values of 
$m_0$ and $m_{1/2}$ determine for the most part the magnitudes of the 
squark and gluino masses, we expect sparticle production rates
in the NUHM1 model to be similar to those in the mSUGRA model for
the same model parameter choices. The reach of the CERN LHC in the case 
of the mSUGRA model has recently been re-evaluated in the
$m_0\ vs.\ m_{1/2}$ plane for various $\tan\beta$ values, and
assuming 100 fb$^{-1}$ of integrated luminosity in Ref.~\cite{lhcreach}.
We display this  reach contour on frame {\it a}) of
Fig.~\ref{fig:nuhm1_col} where as in 
Fig.~\ref{fig:nuhm1_planes}{\it a}) we take
$A_0=0$, $\tan\beta =10$ and $m_\phi =m_0$ (mSUGRA case). 
We also show the WMAP allowed region ($\Omega_{\tz_1}h^2<0.13$) 
as the one shaded in green.
The low
$m_0$ portion of the reach contour extends to $m_{1/2}\sim 1.3$ TeV, and
corresponds roughly to $m_{\tq}\sim m_{\tg}\sim 3$ TeV. The high $m_0$
portion of the reach contour extends to $m_{1/2}\sim 0.7$ TeV, and
corresponds to $m_{\tg}\sim 1.8$ TeV, while squarks are in the multi-TeV
range, and essentially decoupled. 
Note that in this frame the parameter $m_0$ ranges all the way to 10 TeV.

In frame {\it b}), with $m_\phi =-2.5 m_0$, 
a much smaller range of $m_0$ is allowed, and the plot only extends to 
$m_0= 2$ TeV. Since the reach is mainly determined by the values of the 
squark and gluino masses, 
we  adapt the reach contours from 
Ref.~\cite{lhcreach} to this non-mSUGRA case. Here, it is seen that the 
LHC reach covers almost all of the allowed $A$ annihilation funnel.
In frames {\it c}) and {\it d}), we show the cases for $m_\phi =1.1 m_0$
and $1.5 m_0$, respectively. Here, the HB/FP type region re-emerges, but at 
much lower $m_0$ values, as discussed in Sec. \ref{sec:nuhm1_ps}. 
The LHC reach is shown
to cover all of the bulk and stau co-annihilation regions, but only
a part of the higgsino annihilation (HB/FP) region.
\FIGURE{\epsfig{file=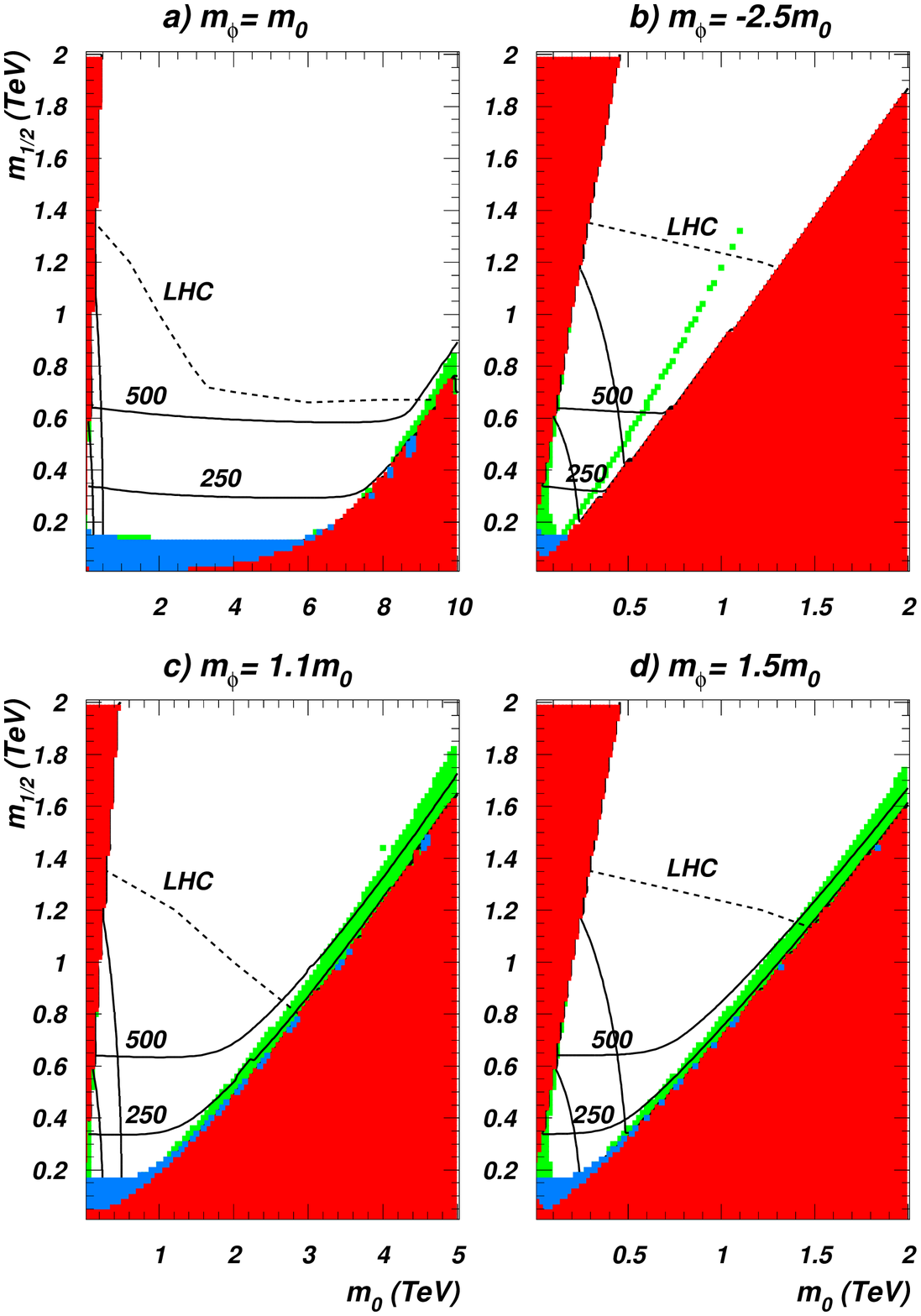,height=14cm,width=14cm}
\caption{Approximate projections
for the reach of the CERN LHC (100 fb$^{-1}$) and ILC
in the NUHM1 model in the $m_0\ vs.\ m_{1/2}$
plane for $\tan\beta=10$, $A_0=0$, $\mu>0$ and $m_t=178$ GeV, for
various choices of $m_\phi /m_0$. The regions shaded in green are
consistent with the WMAP constraint $\Omega h^2 < 0.13$, while those
shaded in red and blue are respectively excluded by theoretical and
experimental constraints discussed in the text. }
\label{fig:nuhm1_col}
}

While we expect a similar reach of the LHC (in terms of $m_{\tq}$ and
$m_{\tg}$ parameters) to be found in both the mSUGRA and NUHM1 models,
the detailed gluino and squark cascade decays will change, as will the
expected SUSY Higgs signals.  To exemplify this, we list in Table
\ref{tab:1} three model points.  The first corresponds to mSUGRA for
$m_0=m_{1/2}=300$ GeV, $A_0=0$, $\tan\beta =10$, $\mu >0$ and $m_t=178$
GeV. We list a variety of sparticle and Higgs boson masses, along with
$\Omega_{\tz_1}h^2$, $BF(b\to s\gamma )$ and $\Delta a_\mu$. The second
and third points listed, NUHM1a and NUHM1b, correspond to the same
mSUGRA model parameters, but with $m_\phi =-735$ GeV and $550$ GeV,
respectively. The mSUGRA case can be seen to have $\Omega_{\tz_1}h^2
=1.2$, and is thus strongly excluded by WMAP data, while the two NUHM1
points have $\Omega_{\tz_1}h^2\sim 0.11$, and give the correct amount of
CDM in the universe. The NUHM1a point has a similar spectrum of
sparticles compared to the mSUGRA case, although the heavier chargino
and neutralinos have increased masses due to the larger value of the
$\mu$ parameter. The main difference is that the heavier Higgs bosons
are relatively light in the NUHM1a case, and can be accessible to LHC
searches as well as at a TeV-scale linear collider. 
In the case of mSUGRA, only
the lightest Higgs $h$ will be detectable at the LHC via direct $h$
production followed by $h\to \gamma\gamma$ decay, or via $t\bar{t}h$ or
$Wh$ production, followed by $h\to b\bar{b}$ decay. The $h$ should also
be observable in the sparticle cascade decays~\cite{spdk2higgs}.  In the
NUHM1a case with $m_A=265$ GeV, the $H$ and $A$ Higgs bosons are much
lighter, and should be detectable via direct $H$ and $A$ production
followed by $H,\ A\to\tau\bar{\tau}$ decay~\cite{lhchiggs}. The reaction
$gb\to tH^+$ followed by $H^+\to\tau^+\nu_\tau$ appears to be on the
edge of observability. If $\tan\beta$ is increased to values beyond 15,
then $H,\ A\to\mu^+\mu^-$~\cite{ckaomumu} should become visible.

\TABLE{
\begin{tabular}{lccc}
\hline
parameter & mSUGRA & NUHM1a & NUHM1b \\
\hline
$m_\phi$ & 300 & -735 & 550 \\
$\mu$ & 409.2 & 754.0 & 180.6  \\
$m_{\tg}$ & 732.9 &  736.2 & 732.0  \\
$m_{\tu_L}$ & 720.9 & 720.5 & 722.4 \\
$m_{\tst_1}$ & 523.4 & 632.4 & 481.0  \\
$m_{\tb_1}$ & 650.0 & 691.6 & 631.0  \\
$m_{\te_L}$ & 364.7 & 366.4 & 364.5  \\
$m_{\te_R}$ & 322.8 & 322.1 & 323.0  \\
$m_{\tw_2}$ & 432.9 & 759.6 & 280.3  \\
$m_{\tw_1}$ & 223.9 & 236.2 & 150.2  \\
$m_{\tz_4}$ & 433.7 & 759.5 & 283.4  \\
$m_{\tz_3}$ & 414.8 & 752.0 & 190.3  \\ 
$m_{\tz_2}$ & 223.7 & 235.8 & 160.7  \\ 
$m_{\tz_1}$ & 117.0 & 118.7 & 102.7  \\ 
$m_A$ & 538.6 & 265.0 & 603.8  \\
$m_{H^+}$ & 548.0 & 278.2 & 613.0  \\
$m_h$ & 115.7 & 116.1 & 115.3 \\
$\Omega_{\tz_1}h^2$& 1.2 & 0.12 & 0.11 \\
$BF(b\to s\gamma)$ & $3.2\times 10^{-4}$ & $4.7\times 10^{-4}$ 
& $2.5\times 10^{-4}$ \\
$\Delta a_\mu    $ & $12.1 \times  10^{-10}$ & $9.4 \times  10^{-10}$ 
& $17.4 \times  10^{-10}$ \\
\hline
\end{tabular}
\caption{\label{tab:1} Masses and parameters in~GeV units
for mSUGRA and two NUHM1 models, where
$m_0=m_{1/2} =300$ GeV, $A_0=0$, $\tan\beta =10$ and $m_t=178$ GeV.
}
\label{tab:nuhm1}
%
}

In the case of NUHM1b, $m_\phi$ is taken large enough that $\mu$ becomes
small, $180.6$ GeV, and the lightest neutralino develops a sufficient
higgsino component to respect the WMAP dark matter constraint. The low
$\mu$ value pulls the various heavier chargino and neutralino masses to
low values ranging from $190-283$ GeV. In this case, gluinos and squarks
will be copiously produced at the LHC. Gluinos dominantly decay via two
body modes to $\tst_1 t$ (BF $\simeq 49$\%) and $\tb_{1,2}b$ (BF $\simeq
39$\%). The lighter stop decays via $\tst_1 \to b\tw_{1,2}$ (53\%),
$\tst_1 \to t\tz_{3}$ (25\%), while $\tb_1 (\tb_2)$ mainly decays to the
two charginos (roughly democratically to all charginos and neutralinos).
The left squarks decay mainly to both charginos and to $\tz_{1,2}$,
while $\tq_R$ mainly decays via $\tq_R \to \tz_{1,2}$. The lighter
chargino decays via three body decays with branching fractions
corresponding to those of the virtual $W$. On the other hand,
$\tz_{2,3}$ decays via three body decays with the leptonic branching
fraction $BF(\tz_{2(3)} \to \ell\bar{\ell}\tz_1) \simeq 1.5(3)$\% per
lepton family.  Finally the heavier chargino mainly decays via $\tw_2
\to W\tz_2, \tw_1 Z$, while $\tz_4 \to \tw_1 W$. It is clear that the
LHC will be awash in SUSY events, with gluino and squark production
being the dominant production mechanism. In this scenario, the total
SUSY cross section is almost $10^4$~fb, so that even at the low
luminosity, we should expect $\sim 100,000$ SUSY events
annually. Moreover, from our discussion of the sparticle decay patterns,
we see that {\it all} the charginos and neutralinos should be accessible
via cascade decays of gluinos and squarks, as envisioned in Ref. \cite{bbkt}. 
It would be extremely
interesting to perform a detailed study of just how much information about
the SUSY spectrum the LHC data would be able to provide in this case.
While detailed simulation would be necessary before definitive
statements can be made, it is plausible that analyses along the lines
carried out in Ref.~\cite{atlas} may yield information about a large part of
the sparticle spectrum, and provide a real connection between collider 
experiments and dark matter searches.

\subsubsection{Linear $e^+e^-$ collider}

The reach of a $\sqrt{s}=0.5$ and 1 TeV international linear $e^+e^-$
collider (ILC) for supersymmetry has been evaluated with special
attention on the HB/FP region in
Ref.~\cite{bbkt_ilc} in the case of the mSUGRA model.  In this study it
was shown that the reach contours in the $m_0\ vs.\ m_{1/2}$ plane are
determined mainly via the reach for sleptons pairs, the reach for
chargino pairs, and partly by the reach in $\tz_1\tz_2$
production. There is also a significant reach for the Higgs bosons $H$,
$A$ and $H^+$ in the large $\tan\beta$ case. The striking  result of
Ref.~\cite{bbkt_ilc} was
that in the WMAP allowed HB/FP region, $|\mu|$ becomes small and charginos
become light, the reach of the ILC extends beyond
that of the LHC. In the HB/FP region of the mSUGRA model, squarks are in
the multi-TeV regime, and effectively not produced at the LHC.  The
signal at the LHC becomes rate limited for $m_{\tg} \agt 1.8$~TeV.

In Fig.~\ref{fig:nuhm1_col}, we also show contours of $m_{\tw_1}$ and
min$(m_{\te_L,\te_R})$ = 250~GeV and 500~GeV: since signals from
chargino and selectron pairs
can be probed at an $e^+e^-$ linear collider nearly up to the
kinematic limit for their production, these contours follow the boundary
of the region that would be probed at the ILC.  Frame {\it a}) shows the
mSUGRA case where an ILC would have an extended reach in the HB/FP
region around $m_0\sim 8-10$ TeV. In frame {\it b}) where the DM-allowed
regions consist only of the $A$-funnel and stau-co-annihilation
corridor, the ILC reach is well below that of the LHC. In frames {\it
c}) and {\it d}), where $m_\phi > m_0$, the HB/FP region has moved to
much lower values of $m_0$. However, in these regions, again $\mu$
becomes small so that charginos become light, and the ILC reach for
$\tw_1^+\tw_1^-$ pairs may exceed the reach of the LHC along the
right-hand edge of parameter space, even in frame {\it d}) where squarks
are comparatively light. 

Concerning the specific SUSY models shown in Table \ref{tab:1}, in the
case of the mSUGRA model, the ILC operating at $\sqrt{s}=500$ GeV would
see of course $Zh$ production, but also $\tw_1^+\tw_1^-$ and
$\tz_1\tz_2$ production. The cross section for the latter process is
$\sim 200$~fb, and since $\tz_2\to\tz_1 Z^0$
essentially all the time, the end points of the energy distribution of $Z^0$
should yield the values 
of $\tz_1$ and $\tz_2$ masses to good precision. This is, of course, 
over and above $m_{\tw_1}$ which can be determined as usual. In the NUHM1a
model, where now the MSSM Higgs sector becomes light, $H^0Z^0$ and
$A^0h$ production will also be possible, allowing a detailed study of the
Higgs sector and possibly a good determination of $\tan\beta$~\cite{bhj}. 
If the ILC energy is increased somewhat above
500 GeV, then $H^+H^-$ also becomes accessible to study.  For the NUHM1b
model, the Higgs bosons again become heavy, but the various heavier
charginos and neutralinos become light. In this case, the final states
$\tz_1\tz_3$, $\tz_1\tz_4$, $\tz_2\tz_2$, $\tz_2\tz_3$, $\tz_2\tz_4$ and
even $\tz_3\tz_4$ as well as $\tw_1^\pm\tw_2^\mp$ are 
kinematically accessible. Thus, a whole host of heavier chargino/neutralino
states would be available for study. If the ILC energy is increased to 
1~TeV, then as in the mSUGRA case, the various slepton pair production as
well as $\tw_1\tw_2$ pair production would
be available for study and SUSY
spectroscopy would become a reality. Moreover,
the heavier Higgs bosons $A$, $H$ and $H^\pm$ which now have
substantial branching fractions to charginos and neutralinos, will also be
accessible to study. 

\section{NUHM2 model}
\label{sec:nuhm2}

\subsection{Overview}

The NUHM2 model is characterized by two additional parameters
beyond the mSUGRA set. The two new parameters may be taken 
to be the GUT scale values of $m_{H_u}^2$ and $m_{H_d}^2$, where
these parameters may take on both positive and negative values.
The model parameter space is given by
\begin{equation}
m_0,\ m_{H_u}^2,\ m_{H_d}^2,\ m_{1/2},\ A_0,\ \tan\beta ,\ {\rm sign}(\mu)\;.
\label{eq:usual}
\end{equation}

We remind the reader that at tree level the Higgs scalar potential is
completely specified by $m_{H_u}^2$, $m_{H_d}^2$, $\mu^2$ and the
parameter $B\mu$. The two minimization conditions allow us to trade
two of $m_{H_u}^2$, $m_{H_d}^2$, $\mu^2$ and $B\mu$ 
in favor of $tan\beta$ and $M_Z^2$, while a third may be traded for
the $CP$ odd Higgs scalar mass $m_A$. In  (\ref{eq:usual}) above, $\mu^2$
and $B\mu$ have been traded for $\tan\beta$ and $M_Z^2$, leaving the
sign of $\mu$ (which enters via the chargino and neutralino mass
matrices) undetermined. Alternatively, 
in the 
NUHM2 model, we could have eliminated $m_{H_u}^2$, $m_{H_d}^2$ and $B\mu$  
leaving $\tan\beta$ together with
the 
weak scale values of $\mu$ and $m_A$ as input parameters. 
Thus, the set
\begin{equation}
m_0,\ \mu ,\ m_A,\ m_{1/2},\ A_0,\ \tan\beta,
\end{equation}
where $\mu$, $m_A$ and $\tan\beta$ are input as weak scale values,
while the remaining parameters are GUT scale values, provides an
alternative parametrization of the NUHM2 model.
In mSUGRA, we have two additional constraints,
$m_{H_u}^2=m_{H_d}^2=m_0^2$, on the scalar potential
and the values of $\mu^2$ and 
$m_A$ are determined.  

We have upgraded Isajet v7.72 to allow not only the input of negative
Higgs squared masses at the GUT scale, but also to accommodate the second
of these parameter sets with weak scale values of $\mu$ and $m_A$ as
inputs, using the non-universal SUGRA (NUSUG) input
parameters~\cite{isajet}.  The NUHM1 and NUHM2 models in Isajet incorporate
REWSB using the RG improved one-loop effective potential,
minimized at an optimal scale $Q=\sqrt{m_{\tst_L}m_{\tst_R}}$ to
account for dominant two-loop contributions.

An important aspect of the NUHM2 model is that RG running of soft masses
is in general modified by the presence of a non-zero $S$ term in
Eq.~(\ref{eq:S}). The quantity $S$, which in the NUHM2 model is given by
$S=m_{H_u}^2-m_{H_d}^2$, enters the third generation soft scalar
squared mass RGEs as,
\begin{eqnarray}
\frac{dm_{Q_3}^2}{dt}&=&{2\over 16\pi^2}\left(-{1\over 15}g_1^2M_1^2-
3g_2^2M_2^2-{16\over 3}g_3^2M_3^2+{1\over 10}g_1^2S+
f_t^2X_t+f_b^2X_b\right),\nonumber \\
&& \label{rgeq1}\\
\frac{dm_{\tst_R}^2}{dt}&=&{2\over 16\pi^2}\left(-{16\over 15}g_1^2M_1^2-
{16\over 3}g_3^2M_3^2-{2\over 5}g_1^2S+2f_t^2X_t\right), \\
\frac{dm_{\tb_R}^2}{dt}&=&{2\over 16\pi^2}\left(-{4\over 15}g_1^2M_1^2-
{16\over 3}g_3^2M_3^2+{1\over 5}g_1^2S+2f_b^2X_b\right), \\
\frac{dm_{L_3}^2}{dt}&=&
{2\over 16\pi^2}\left(-{3\over 5}g_1^2M_1^2-
3g_2^2M_2^2-{3\over 10}g_1^2S+f_\tau^2X_\tau\right), \\
\frac{dm_{\ttau_R}^2}{dt}&=&{2\over 16\pi^2}\left(-{12\over 5}g_1^2M_1^2+
{3\over 5}g_1^2S+2f_\tau^2X_\tau\right) . \label{rgeq4}
\end{eqnarray}
The first and second generation soft mass RGEs are similar, but with
negligible Yukawa coupling
contributions. The Higgs boson soft mass RGEs are as given by
(\ref{eq:mhu}) and (\ref{eq:mhd}).  The coefficients of the $S$ terms are
all proportional to the weak hypercharge assignments, so that this term
provides a source of intra-generational mass splitting.  When $S$ is
large and positive ({\it i.e.}, when $m_{H_u}^2>m_{H_d}^2$), 
the mass parameters for 
$\ttau_R,\ \te_R$ and $\tmu_R$ are the most suppressed, 
while those for $\tq_R$ and $\tell_L$ are enhanced.
If $S$ is large and negative, the situation is exactly reversed. For
large values of $|S|$, the sfermion mass ordering as well as
mixing patterns of third generation sfermions
may be altered from mSUGRA expectations, or for that matter expectations
in many other models of sparticle masses. For instance, it is 
possible that $m_{\tell_L}< m_{\tell_R}$; moreover, while the lighter
stau is usually expected to be dominantly $\ttau_R$ in most models, this may
no longer be the case in the NUHM2 model. 

As a simple illustration of how the spectrum of the NUHM2 model varies with
the Higgs boson mass parameters,
in Fig.~\ref{fig:nuhm2_mass2} we show the physical masses of various
sparticles versus $\Delta m_H\equiv m_0-sign(m_{H_u}^2)\cdot\sqrt{|m_{H_u}^2|} 
= sign(m_{H_d}^2)\cdot\sqrt{|m_{H_d}^2|}-m_0$.
This is a one parameter section of the NUHM2 parameter space we call
the Higgs splitting (HS) model, with $\Delta m_H=0$
corresponding to the mSUGRA model.
Large positive $\Delta m_H$ gives rise to a large {\it negative} $S$,
and vice versa. 
In our example, 
we take $m_0=m_{1/2}=300$ GeV, with $A_0=0$, $\tan\beta
=10$ and $\mu >0$.  As $\Delta m_H$ increases, we see that in the first
generation, the $\te_R,\ \td_R$ and $\tu_L$ masses all increase, while
$\te_L$ and $\tu_R$ masses decrease. In mSUGRA, $m_{\te_R}$ is always
less than $m_{\te_L}$; in NUHM2 models, this mass ordering may be
reversed.  At the highest allowed values of $\Delta m_H$, the
$\tnu_\tau$ and $\ttau_1$ mass values become light (and $\ttau_1$ is then
dominantly $\ttau_L$), enhancing
$t$-channel $\tz_1\tz_1\to \nu_\tau\bar{\nu}_\tau,\ e^+e^-,\
\nu_\mu\bar{\nu}_\mu,\ \mu^+\mu^-$ annihilation in the early universe,
which lowers the relic density to within the WMAP bound.  When
$m_{\tnu_{\tau}},\ m_{\ttau_1}\simeq m_{\tz_1}$, then co-annihilation
reduces the relic density even further.

\FIGURE{\epsfig{file=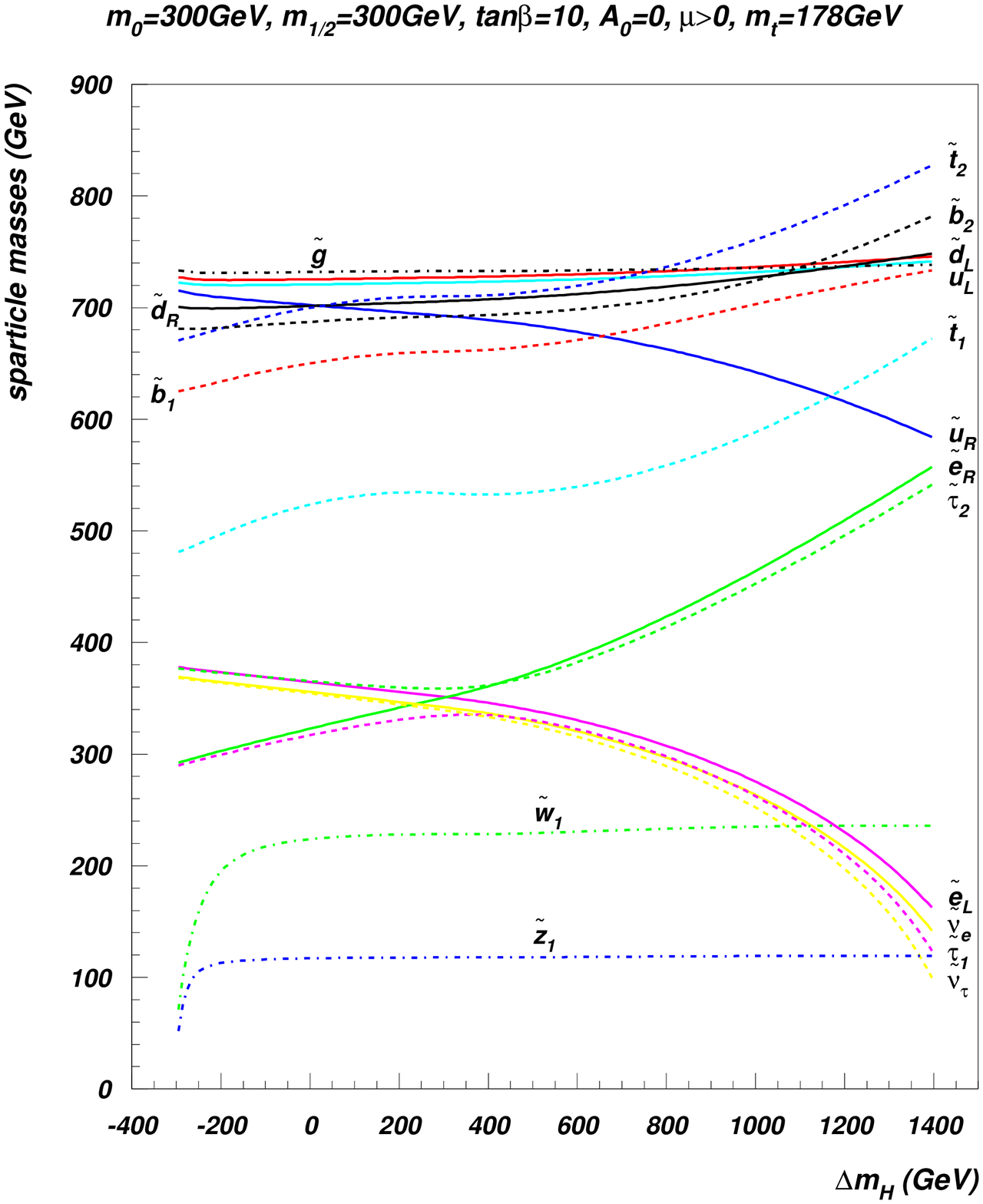,width=14cm}
\caption{Variation in sparticle masses versus $\Delta m_H$
in the NUHM2 model, for $m_0=300$ GeV,
$m_{1/2}=300$ GeV, $A_0=0$, $\tan\beta =10$, $\mu>0$ and $m_t=178$ GeV.
}
\label{fig:nuhm2_mass2}
}
\FIGURE{\epsfig{file=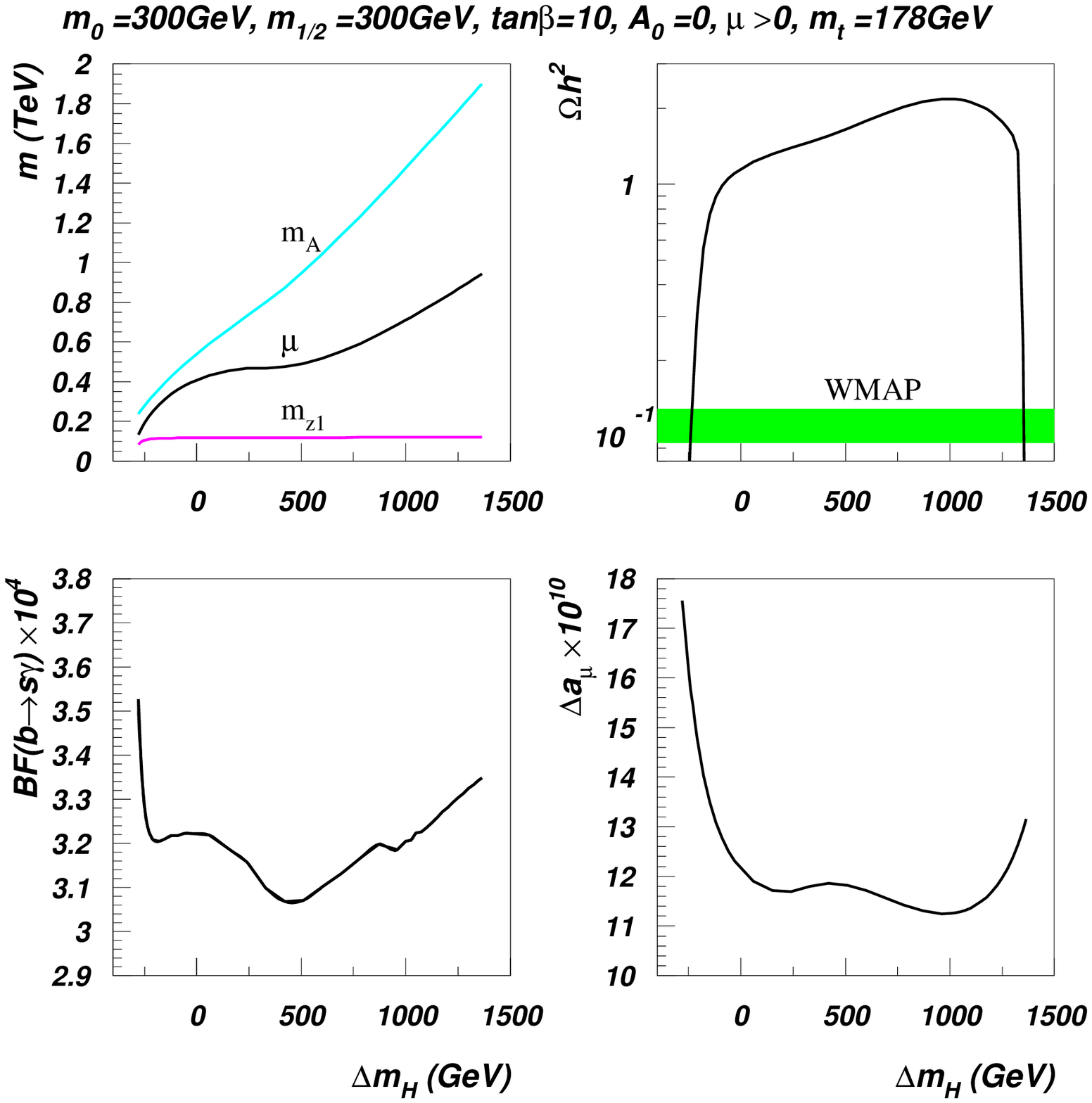,width=14cm}
\caption{Variation in {\it a}) $m_A$, $\mu$ and $m_{\tz_1}$, {\it b})
$\Omega_{\tz_1}h^2$, {\it c}) $BF(b\to s\gamma )$ and {\it d}) $\Delta
a_\mu$ versus $\Delta m_H\equiv (m_{H_d}-m_{H_u})/2$ in the NUHM2 model,
for $m_0=300$ GeV, $m_{1/2}=300$ GeV, $A_0=0$, $\tan\beta =10$, $\mu>0$
and $m_t=178$ GeV.}
\label{fig:nuhm2_rates2}
}

Some aspects of NUHM2 phenomenology as a function of $\Delta m_H$ are
illustrated in Fig.~\ref{fig:nuhm2_rates2}.  In frame {\it a}) we show
the values of $\mu$, $m_A$ and $m_{\tz_1}$ versus $\Delta m_H$. For
negative values of $\Delta m_H$, both $\mu$ and $m_A$ are small,
and we have a region of higgsino and (possibly) $A$-funnel annihilation. For
$\Delta m_H$ large and positive, the relic density, shown in frame {\it
b}), drops because left sleptons and sneutrinos become very light. The
value of $BF(b\to s\gamma )$ and $\Delta a_\mu$ are also shown in frames
{\it c}) and {\it d}). These rise for negative values of $\Delta m_H$
because the charged Higgs bosons and the lighter charginos and
neutralinos become very light with the -inos developing 
significant higgsino components.

In Fig.~\ref{fig:nuhm2_mass1} we show again the variation in sparticle
masses with $\Delta m_H$, but this time for $m_0=1450$ GeV,
$m_{1/2}$=300$ GeV, A_0=0$, $\tan\beta =10$ and $\mu >0$.  In this case,
the large scalar masses yield a large $S$ term in the RGEs, and the
hypercharge enhancement/suppression is accentuated.  We see, as noted in
Ref.~\cite{auto2} for Yukawa unified models, that the $\tu_R$ and
$\tc_R$ squarks are driven to very low mass values as $\Delta m_H$
increases.  At the high end of the $\Delta m_H$ range, they become the
lightest squarks, even lighter than the $\tst_1$. The large $\Delta m_H$
parameter space ends when $\tu_R$ becomes a charged/colored LSP, in
violation of restrictions forbidding such cosmological relics.  We also
see that as $\Delta m_H$ increases, the $\tst_1$ mass at first
increases, then decreases, then increases again. The initial increase is
because as $\Delta m_H$ increases, $X_t$ decreases, leading to reduced
Yukawa coupling suppression of the soft SUSY breaking mass parameters of
top squarks.  For still larger values of $\Delta m_H$, the $S$ term
grows and leads to a suppression of the $\tst_R$, and hence $\tst_1$
mass. Finally, as $\Delta m_H$ is increased even more, the $X_t$ term
becomes large and {\it negative}, and again resulting in an increase in
the top squark soft masses.  We have checked that throughout this range
of $\Delta m_H$, the lighter top squark remains predominantly
right-handed.
\FIGURE{\epsfig{file=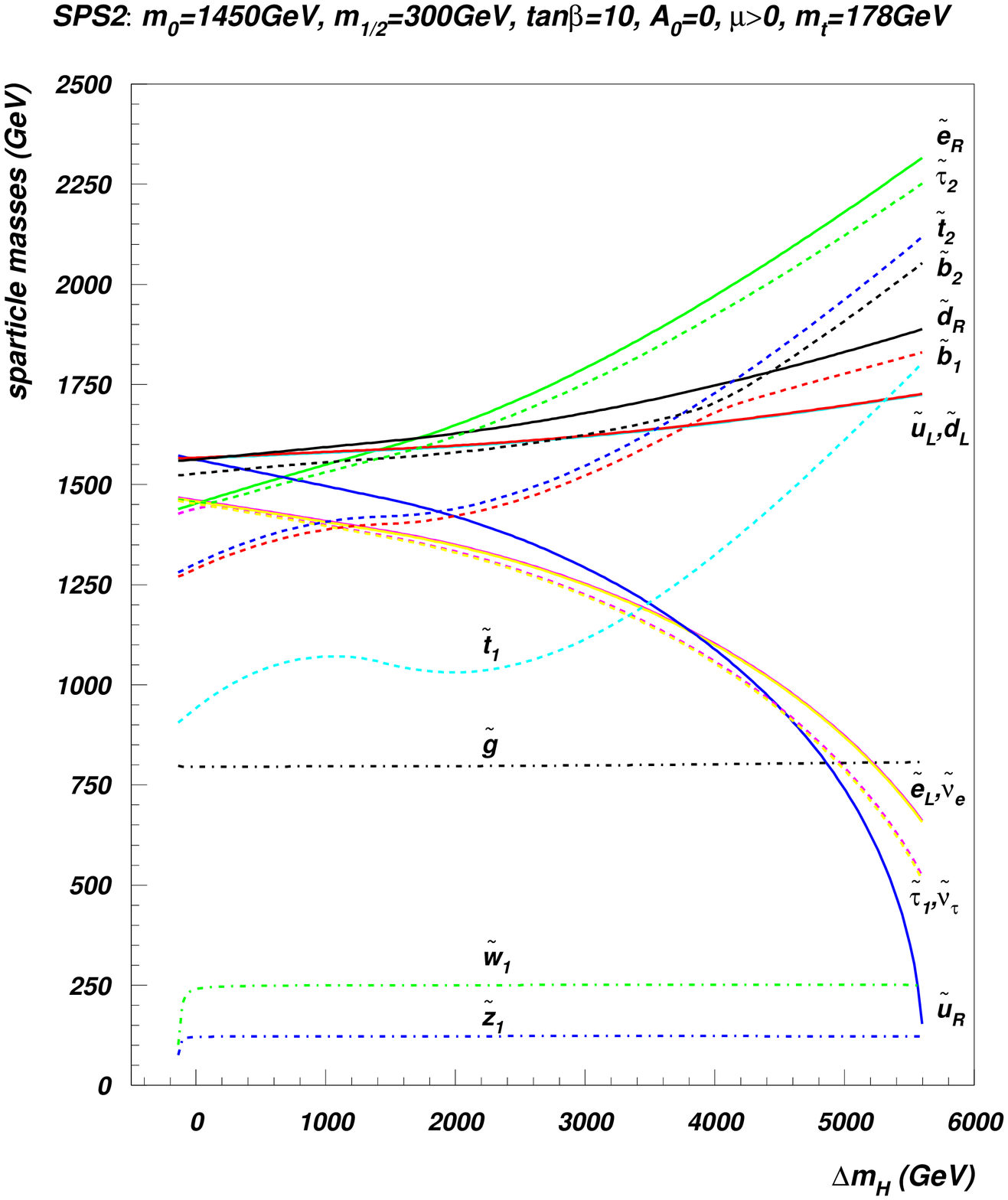,width=14cm}
\caption{Variation of sparticle masses versus $\Delta m_H$ defined
in the text for the NUHM2 model, with $m_0=1450$ GeV,
$m_{1/2}=300$ GeV, $A_0=0$, $\tan\beta =10$, $\mu>0$ and $m_t=178$ GeV.}
\label{fig:nuhm2_mass1}
}
\FIGURE{\epsfig{file=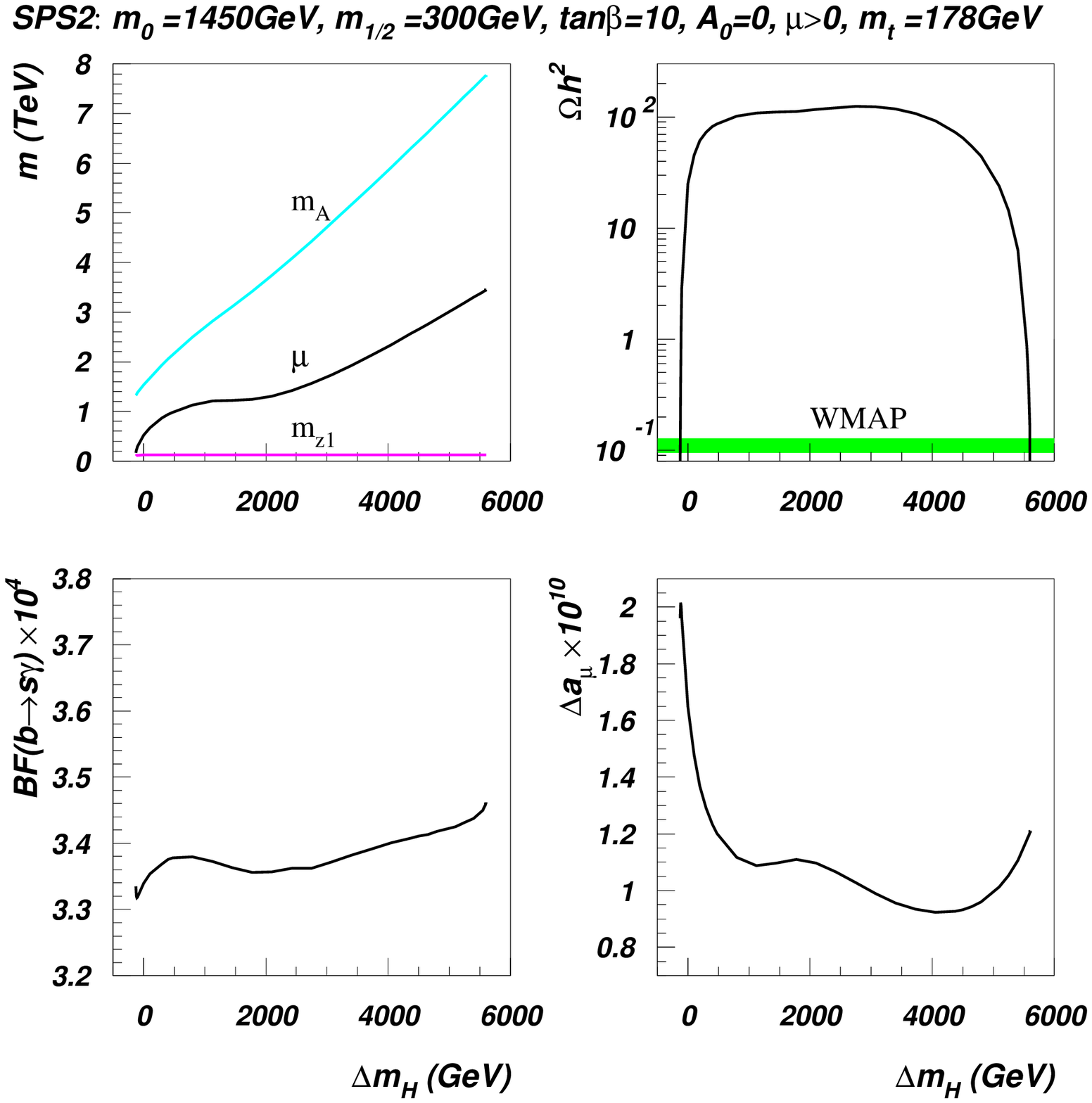,width=14cm}
\caption{Variation in {\it a}) $m_A$, $\mu$ and $m_{\tz_1}$, {\it b})
$\Omega_{\tz_1}h^2$, {\it c}) $BF(b\to s\gamma )$ and {\it d}) $\Delta
a_\mu$ versus $\Delta m_H\equiv (m_{H_d}-m_{H_u})/2$ in the NUHM2 model,
for $m_0=1450$ GeV, $m_{1/2}=300$ GeV, $A_0=0$, $\tan\beta =10$, $\mu>0$
and $m_t=178$ GeV.}
\label{fig:nuhm2_rates1}
}

In Fig.~\ref{fig:nuhm2_rates1}, we show the same frames as in
Fig.~\ref{fig:nuhm2_rates2}, but now for the $m_0=1450$ GeV case. At low
$\Delta m_H$ values, again $\mu$ gets to be small, so that the neutralino
becomes higgsino-like leading to efficient annihilation in the early
universe. At very
large $\Delta m_H$ values, the relic density is again in accord with the
WMAP analysis -- this time 
because the squarks become so light that neutralinos can
efficiently annihilate via $\tz_1\tz_1\to u\bar{u},\ c\bar{c}$ processes
occurring via $t$-channel $\tu_R$ and $\tc_R$ exchange. Furthermore, if
$m_{\tu_R}$ and $m_{\tc_R}$ are in the 100-200 GeV range, they may be
accessible to Tevatron searches!
The light squarks also lead to a greatly enhanced neutralino-proton
scattering rate, and hence to large rates for direct detection of relic
neutralinos~\cite{auto2}. 
There is
a small enhancement in $\Delta a_\mu$ at low $\Delta m_H$ where 
charginos and neutralinos become light and higgsino-like, 
leading to larger
chargino-sneutrino and neutralino-smuon loop contributions
in the evaluation of $(g-2)_\mu$~\cite{bbft}.

\subsection{NUHM2 model: parameter space}

Our first display of the parameter space of the NUHM2 model is in
Fig.~\ref{fig:nuhm2_mhumhdplane}, where we show the $sign(
m_{H_u}^2)\cdot\sqrt{|m_{H_u}^2|}\ vs.  \ sign(
m_{H_d}^2\cdot\sqrt{|m_{H_d}^2|}$ plane for $m_0=m_{1/2}=300$ GeV, with
$A_0 =0$, $\tan\beta =10$, $\mu >0$ and $m_t=178$ GeV.  In frame {\it
a}) we show the allowed parameter space as the white region, while
theoretically excluded parameter choices are red. The region to the
right is excluded because $\mu^2 < 0$. In the red region at the bottom,
$m_A^2 < 0$, while in that on the top, $\tz_1$ is not  the LSP. 
\FIGURE{\epsfig{file=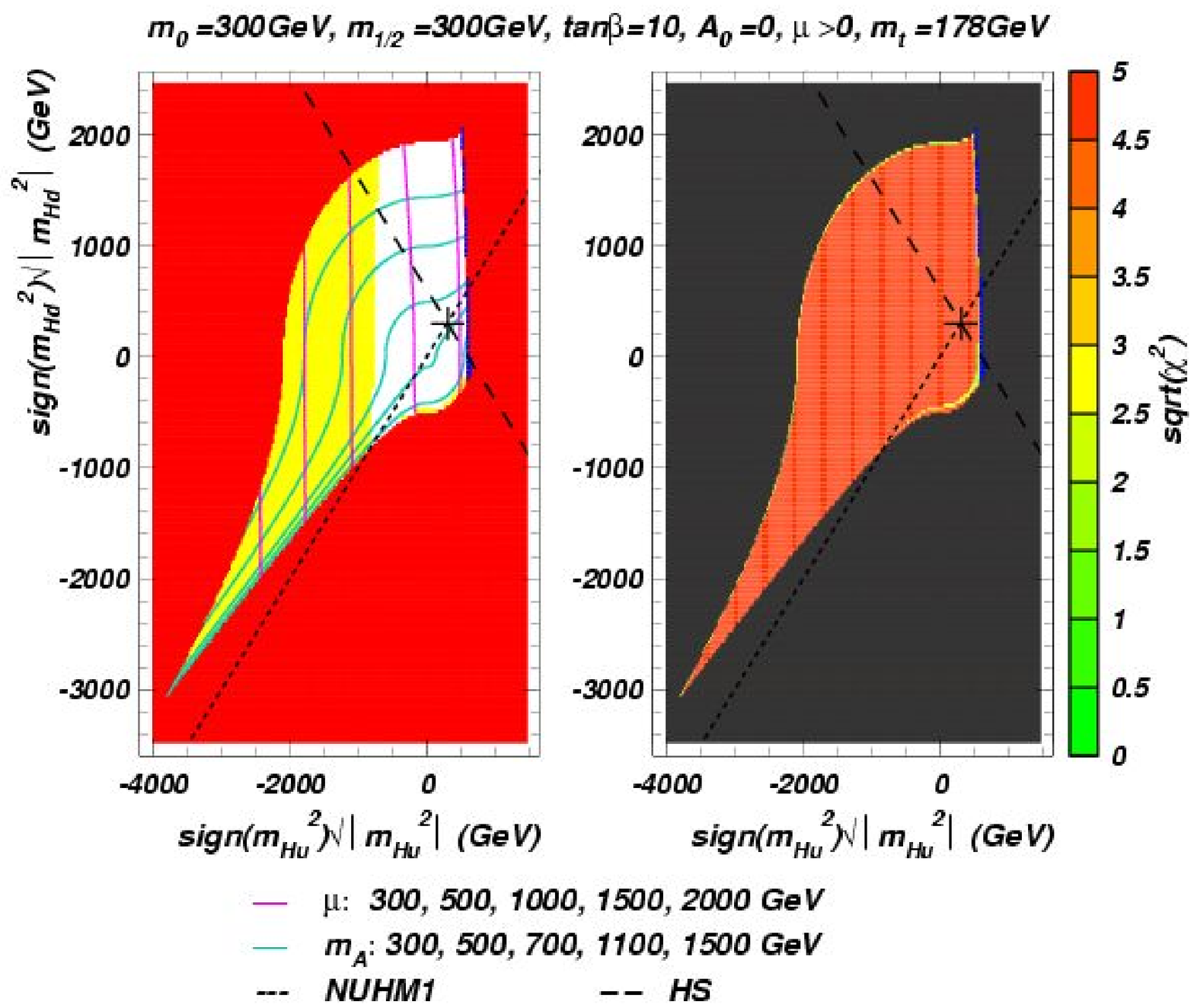,width=14cm}
\vspace*{-0.8cm}
\caption{Plot of allowed parameter space in the $m_{H_u}\ vs.\ m_{H_d}$
plane of the NUHM2 model for for $m_0=m_{1/2}=300$ GeV, $A_0=0$,
$\tan\beta =10$, $\mu >0$ and $m_t=178$ GeV. In frame {\it a}), we show
contours of $\mu$ and $m_A$, while in frame {\it b}) we show values of
$\chi^2$. The yellow region in frame {\it a}) is where so-called ``GUT
stability bound'' is violated. The short-dashed black line denotes the
parameter space of the NUHM1 model. The cross denotes the mSUGRA model,
while the long-dashed line gives the model where the Higgs scalar mass
parameters are split as in Fig.~\ref{fig:nuhm2_mass2}.
The blue region is excluded by LEP2.
}
\label{fig:nuhm2_mhumhdplane}
}
The blue region is
allowed theoretically, but here, $|\mu |$ dives to small values
yielding $m_{\tw_1}<103.5$, in violation of the
bound from LEP2.  
The parameter space of the NUHM1 model is
shown by the black dashed line, where $m_{H_u}^2=m_{H_d}^2$, while
the mSUGRA value point where $m_{H_u}^2=m_{H_d}^2=m_0^2$ is shown by a
black cross. The reader will notice that while the 
bulk of the parameter space of
the NUHM2 model lies {\it above} this dashed black line where
$m_{H_d}^2 > m_{H_u}^2$, there is a small portion for small values of
$|m_{H_{u,d}}^2|$ values this is not the case. The reason for this
asymmetry is that as seen from the EWSB conditions (\ref{eq:ewsb1}) and
(\ref{eq:ewsb2}), the {\it  weak scale} values of the Higgs mass squared
parameters must satisfy $m_{H_d}^2 > m_{H_u}^2$, with $m_{H_u}^2 <
0$: if the former inequality is badly violated at the GUT scale,
radiative corrections cannot ``correct this'', and the correct
pattern of EWSB is not obtained.
We also show contours of $\mu$ (magenta) ranging from
300-2000 GeV, where 300 GeV contour is on the far right-hand
side. Contours of $m_A$ ranging from 300-1500 GeV are also shown,
increasing from bottom to top.  
The $\sqrt{\chi^2}$ value is shown in frame {\it b}), which
shows most of the parameter space is excluded. The exception is the
narrow green/yellow region near the lower edge of allowed parameter
space, which is the $A$-funnel, and on the right-most edge of parameter
space, barely visible, is the higgsino region.
The narrow green region at the upper boundary of parameter space
corresponds to the slepton (or squark for large $m_0$) 
co-annihilation region.
\FIGURE{\epsfig{file=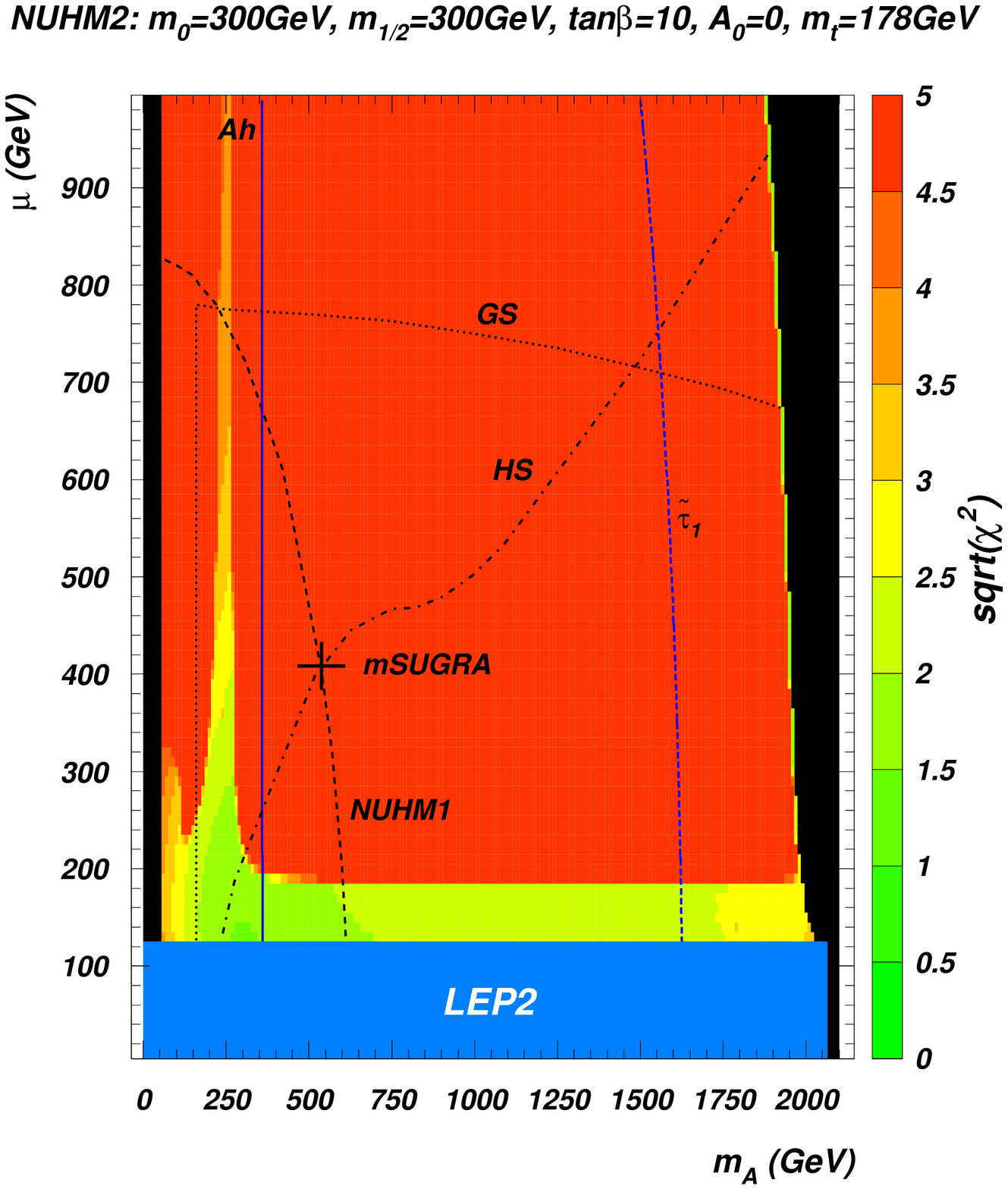,width=12cm,height=11cm}
\caption{Plot of regions of $\sqrt{\chi^2}$ in the $\mu\ {\rm vs.} \ m_A$
plane for $m_0=m_{1/2}=300$ GeV, 
$A_0=0$, $\tan\beta =10$ and $m_t=178$ GeV for $\mu >0$. The line
labeled HS denotes the NUMH2 model where just the Higgs mass parameters
are split as in Fig.~\ref{fig:nuhm2_mass2}.
The region to the left of the $Ah$ contour is where $Ah$ production is 
accessible to a $\sqrt{s}=500$ GeV ILC, while the region to the right of
the $\ttau_1$ contour is accessible to ILC via stau pair searches.
}
\label{fig:nuhm2_chi2}
}

The DM allowed regions of parameter space show up more prominently if
the parameters $m_{H_u}^2$ and $m_{H_d}^2$ are traded for $m_A$ and
$\mu$ as inputs. We display in Fig.~\ref{fig:nuhm2_chi2} the
$\sqrt{\chi^2}$ values for the same parameter space as in
Fig.~\ref{fig:nuhm2_mhumhdplane}, but this time in the $m_A\ vs.\ \mu$
plane. Again most of the parameter space is excluded, although in this
mapping the higgsino region shows up as the broad band of green/yellow
at low $\mu$ values, while the $A$-annihilation funnel shows up as the
vertical band running upwards near $m_A\sim 250 $ GeV.  This plot
highlights the importance of the measure of parameter space when
deciding the likelihood that any particular framework satisfies some
empirical (or, for that matter, theoretical) criteria: the tiny
green/yellow sliver along the right edge in
Fig.~\ref{fig:nuhm2_mhumhdplane} is expanded into the band, while the
thin sliver at the bottom shows up as the $A$ funnel. Notice also the
thin green/yellow region for very large $m_A$ values, where the
existence of light sleptons brings the relic density 
prediction into accord with
the WMAP value of $\Omega_{CDM}h^2$.
\FIGURE{\epsfig{file=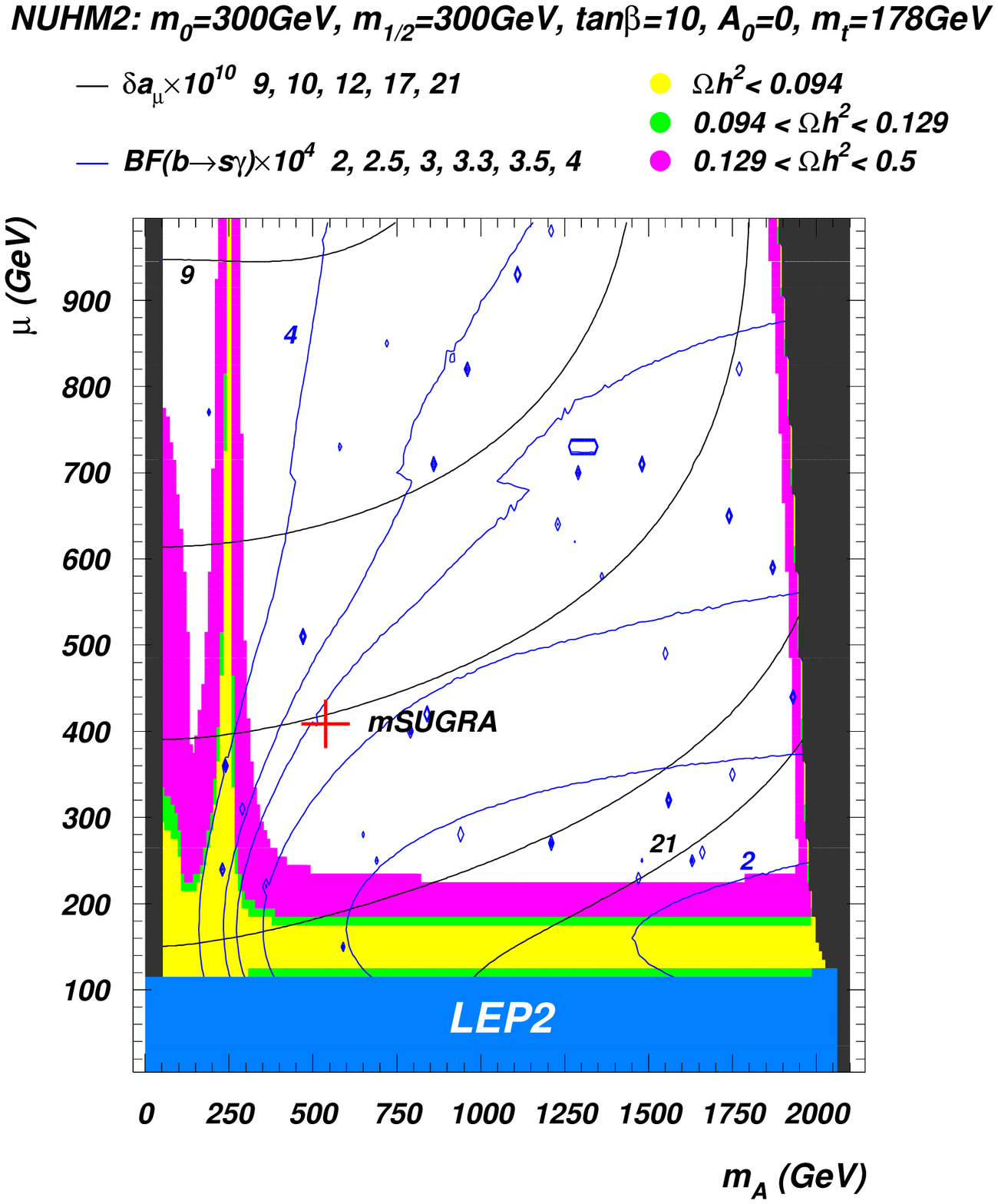,width=7cm}
\epsfig{file=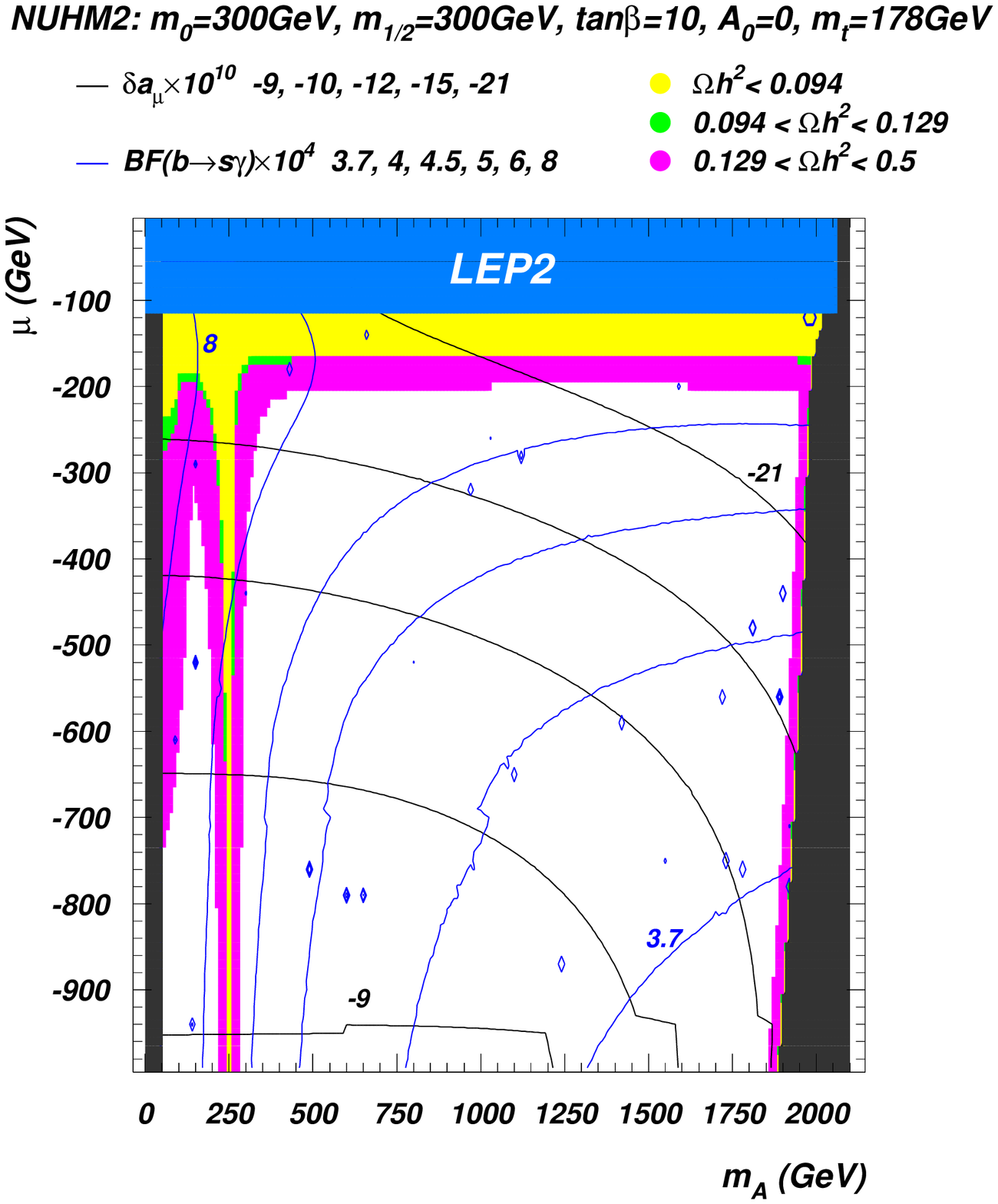,width=7cm}
\vspace*{-0.5cm}
\caption{Ranges of $\Omega_{\tz_1}h^2$ together with contours  of
$BF(b\to s\gamma )$ and $\Delta a_\mu$ in the $\mu\ vs.\ m_A$
plane for $m_0=m_{1/2}=300$ GeV, 
$A_0=0$, $\tan\beta =10$ and $m_t=178$ GeV. For very large values of
$m_A$, we have the stau co-annihilation region. 
In frame {\it a}), we show contours for $\mu <0$ and in
frame {\it b}) we show contours for $\mu >0$.
}
\label{fig:nuhm2_contours}
} 
The NUHM1 model extends along the black dashed contour arc, with the
mSUGRA model denoted by a cross.  The regions away from the NUHM1 model
arc denote SUSY mass spectra which are phenomenologically different from
either the mSUGRA or NUHM1 model. In fact the points with low $\mu$ {\it
and} low $m_A$, which only occur in the NUHM2 model, are somewhat
favored by the combined constraints.  The regions above and left of the
GS contour violate the GS condition, while the HS contour denotes the
path of the HS model through the NUHM2 model parameter space.

The predictions for $\Omega_{\tz_1}h^2$ and contours
for $BF(b\to s\gamma )$ and $\Delta a_\mu$ are separately shown in
Fig.~\ref{fig:nuhm2_contours} for {\it a}) $\mu >0$ and also for the
disfavored value {\it b}) $\mu <0$.  As before, we take
$m_0=m_{1/2}=300$ GeV, $A_0=0$, $\tan\beta =10$ and $m_t=178$ GeV.  In
frame {\it a}), the bulk of the allowed parameter space is determined by
the WMAP allowed region. Within this region, the values of $BF(b\to
s\gamma )$ and $\Delta a_\mu $ determine the best fit, which turns out
to be $m_A\sim 300$ GeV and $\mu \sim 130$ GeV. In this region,
charginos and neutralinos as well as MSSM Higgs bosons are all
relatively light. In frame {\it b}), for $\mu <0$, it is seen that
$BF(b\to s\gamma )$ is $\sim 4\times 10^{-4}$ in the low right-hand
region. For smaller values of $|\mu |$ and $m_A$, the value of $BF(b\to
s\gamma )$ only increases, pushing the $\chi^2$ to large values all
over the WMAP allowed region.

\subsection{Dark matter detection: the NUHM2 model}

We show in Fig.~\ref{fig:dm_NUHM2_b} and \ref{fig:dm_NUHM2_a} the reach 
contours for the various detection channels introduced in Sec.~\ref{sec:dm}, 
respectively for the Burkert Halo Model (Fig.~\ref{fig:dm_NUHM2_b}) 
and for the Adiabatically Contracted N03 Halo 
Model (Fig.~\ref{fig:dm_NUHM2_a}): parameter space points lying below, 
or to the left, of the reach contours will yield detectable signals
via the corresponding
searches. The dashed black lines mark the locus of 
points appropriate to the one-parameter NUHM1 model, where 
$m_{H_u}^2=m_{H_d}^2$ at $Q=M_{GUT}$.
The red cross indicates the 
particular point given by the universal mSUGRA case.
\FIGURE{\epsfig{file=dm_NUHM2_b.eps,width=15cm}
\caption{The reach contours ({\em i.e.} the iso-VR=1 lines) for 
various dark matter detection techniques in the $(m_A,\mu)$ plane. 
Points lying to the bottom-left of the lines are within reach of the 
future experimental facilities, as described in Sec.~\ref{sec:dm}. 
The black dashed line indicates the parameter space of the NUHM1 model 
in this plane, while the red cross locates the mSUGRA universal case. 
}
\label{fig:dm_NUHM2_b}
}
%
\FIGURE{\epsfig{file=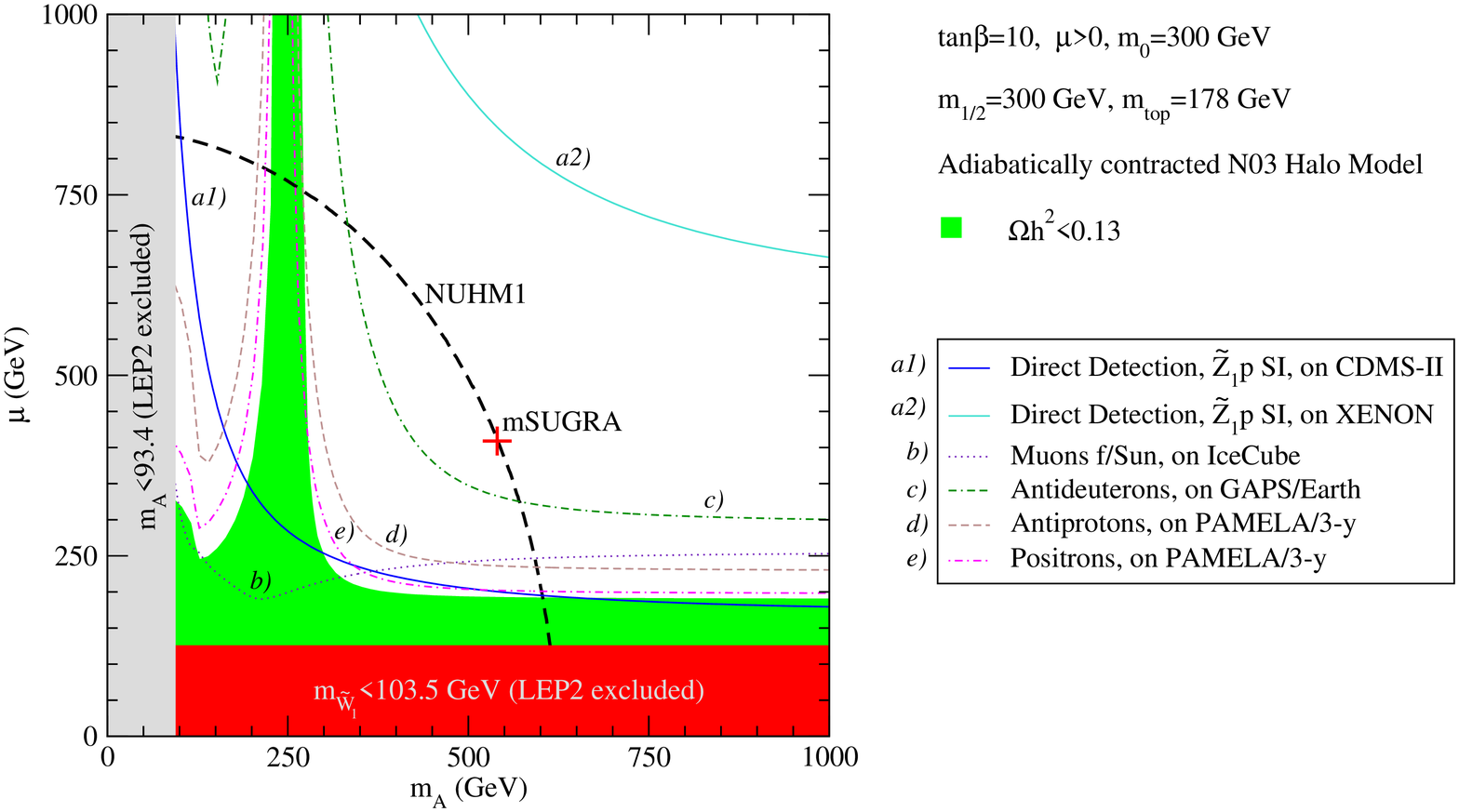,width=15cm}
\caption{The same as in Fig.~\ref{fig:dm_NUHM2_b}, but for the Adiabatically 
contracted N03 Halo Model. The VR contour for gamma rays from the 
galactic center is not shown, since the whole parameter space range 
shown in the figure features a VR larger than 1 in that detection 
channel (see the discussion in the text).}
\label{fig:dm_NUHM2_a}
}
%

For the chosen values of mSUGRA parameters (which include gluino and
squark masses up to several hundred GeV), stage-3 direct detection
experiments will probe the bulk of the $\mu-m_A$ plane allowed by
cosmology, independently of the halo model under consideration. 
The exception is the region with very large values of $m_A\sim 1.8$~TeV
(not shown in the figure) where the relic density is in accord with the
WMAP observation because the sleptons become very light. 
A large
portion of the WMAP allowed region will also be within reach of stage-2
detectors, particularly for low values of the $\mu$ parameter. The shape
of the VR curves is again readily understood in terms of the interplay
of the two effects that we discussed in detail in Sec.~\ref{sec:dm1}:
the enhancement of the heavy $CP$-even Higgs exchange channel at low
$m_A$, and the increased higgsino fraction, at low values of the $\mu$
parameter.

The expected flux of muons from neutralino annihilations in the Sun 
is particularly sensitive to the value of the $\mu$ parameter: 
if the latter is sufficiently low, it provides a large enough 
spin-dependent neutralino-proton scattering cross section and a 
large capture rate of neutralinos in the core of the Sun, hence 
giving a large enough signal at IceCube.

A comparison of Figs.~\ref{fig:dm_NUHM2_b} and \ref{fig:dm_NUHM2_a}
shows what we had alluded to in the last section: the Burkert halo
profile yields conservative predictions for the prospects for detection
of relic LSPs. 
Indirect detection rates essentially track the size of the mass-rescaled
pair annihilation rate $\langle \sigma v\rangle/m^2_{\widetilde
Z_1}$. 
The difference between the profiles is especially accentuated for 
gamma rays, where the five orders of magnitude enhancement mentioned in
Sec.~\ref{sec:dm} implies that the entire plane will be covered for
the Adiabatically Contracted Halo Model.
The difference between the models is also considerable 
for the various antimatter searches, where once again, the Burkert
profile leads to the most conservative prediction.

%
%
\FIGURE{\epsfig{file=dd_DELTAMH_1.eps,width=12cm}
\caption{The neutralino-proton spin-independent scattering cross section
$\sigma_{\widetilde{Z_1}p}^{SI}$ as a function of $\Delta m_H$ (black
solid line), at $m_0=300$ GeV. The other input parameters are specified
in the figure. The dotted red line indicates the current 90\% CL
exclusion limit
on $\sigma_{\widetilde{Z_1}p}^{SI}$ delivered by the CDMS
experiment~\cite{Akerib:2004fq}, while the blue dashed lines and the
green dot-dashed lines correspond to the projected 90\%CL exclusion
limits for Stage-2 and Stage-3 detectors, taking as benchmark
experiments the next-generation CDMS-II detector and the XENON-1~ton
facility.
}\label{fig:dd_DELTAMH_1}
}
\FIGURE{\epsfig{file=dd_DELTAMH_2.eps,width=12cm}
\caption{Same as in Fig.\ref{fig:dd_DELTAMH_1}, but at $m_0=1450$
GeV.}\label{fig:dd_DELTAMH_2}
}

We show in Fig.~\ref{fig:dd_DELTAMH_1} ($m_0=300$ GeV) and
Fig.~\ref{fig:dd_DELTAMH_2} ($m_0=1450$ GeV) the neutralino-proton
spin-independent scattering cross section
$\sigma_{\widetilde{Z_1}p}^{SI}$ as a function of $\Delta m_H$, together
with the current experimental limit and the projected sensitivity of
Stage-2 and Stage-3 detectors. In both cases,
$\sigma_{\widetilde{Z_1}p}^{SI}$ increases significantly for large
negative values of $\Delta m_H$, in the region where the neutralino gets
a large higgsino fraction since $|\mu|$ becomes small (left
panels). Stage-3 direct detection experiments will thoroughly probe the
WMAP allowed region in the left part of the plots, and Stage-2 detectors
will have access to a wide portion of it. As for the large positive end
of the $\Delta m_H$ range, a WMAP compatible relic abundance is
achieved, in both cases, through sfermion co-annihilations. In the
$m_0=300$ GeV case squarks are heavy, and a light slepton sector does
not particularly facilitate direct detection of the neutralino, which is
found to lie even beyond Stage-3 detectors.
\footnote{Although we can
envision that the neutralino electron scattering may be resonantly
enhanced if $m_{\te_L}\simeq m_{\tz_1}$, the energy transferred to the
electron is very small because the electron is very light. In this
respect, the situation is different for the case of scattering off a
nucleon discussed below.}
In the
$m_0=1450$ GeV case, the co-annihilating partners 
participate in the neutralino-proton
scattering through $s$-channel squark exchange diagrams. This results in
resonant squark contributions to neutralino nucleon scattering which
enormously enhance $\sigma_{\widetilde{Z_1}p}^{SI}$ when $m_{\tilde
q}\rightarrow m_{\widetilde{Z_1}}$. The steepness of
$\sigma_{\widetilde{Z_1}p}^{SI}$ as a function of $\Delta m_H$, in the
$m_0=1450$ GeV case, is further increased by the occurrence of an
accidental cancellation between the up-squark neutralino-quark amplitude
and the charm-squark gluon-mediated amplitude at $\Delta m_H\simeq 5530$
GeV.

\FIGURE{\epsfig{file=nt_DELTAMH_1.eps,width=12cm}
\caption{The flux of muons from the Sun generated by charged current
interactions of neutrinos produced in the center of the celestial body
by neutralino-pair annihilations, integrated above a 1 GeV threshold
(solid black line), as a function of $\Delta m_H$ (black solid line),
at $m_0=300$ GeV, with other parameters as specified in the
figure. The red dotted curves indicate the current 90\% CL Super-Kamiokande
limits on the muon flux from the Sun, while the dashed blue lines the
expected 90\% CL IceCube sensitivity. We corrected for the energy threshold
mismatch in the computation of the projected IceCube exclusion curve,
and we used a hard neutrino spectrum in the left panel and a soft
neutrino spectrum in the right panel (see the text for
details).}\label{fig:nt_DELTAMH_1}
}
\FIGURE{\epsfig{file=nt_DELTAMH_2.eps,width=12cm}
\caption{Same as in Fig.~\ref{fig:nt_DELTAMH_1}, but at $m_0=1450$
GeV.}\label{fig:nt_DELTAMH_2}
}

Fig.~\ref{fig:nt_DELTAMH_1} ($m_0=300$ GeV) and Fig.~\ref{fig:nt_DELTAMH_2}
($m_0=1450$ GeV) show our results for the expected muon flux from the
sun for the supersymmetric cases that we have been examining. As for
the neutralino-proton spin-independent scattering cross section
discussed above, the increase in the higgsino fraction in the left
part of the left panels dictates a larger neutralino pair annihilation
cross section as well as a larger spin-dependent neutralino-proton
scattering cross section. As a result, when the neutralino thermal
relic abundance enters the WMAP region, the expected muon flux gets
close to or above the IceCube projected sensitivity for a hard neutrino
spectrum, and even exceed the Super-Kamiokande limit in the $m_0=1450$
GeV case. The abrupt drop in the muon flux visible in the $m_0=1450$
GeV figure, close to the LEP-2 chargino mass limit, is due to the
gauge bosons thresholds. As regards the sfermion co-annihilation
regions (right panels), light sleptons induce only a slightly larger
neutralino pair annihilation cross section (Fig.~\ref{fig:nt_DELTAMH_1}),
while the squark-mediated resonant increase in
$\sigma_{\widetilde{Z_1}p}$ produces an extremely large muon flux in
the $m_0=1450$ GeV scenario (Fig.~\ref{fig:nt_DELTAMH_2}).

%
\FIGURE{\epsfig{file=dm_DELTAMH_1.eps,width=16cm}
\caption{The Dark Matter detection Visibility Ratios (VR) for various
detection strategies, as a function of the GUT-scale Higgs mass
splitting $\Delta m_H$, at $m_0=300$ GeV. The other input parameters
are specified in the figure. When the VR is above the ``Visibility
Threshold'' ({\em i.e.} the VR=1 line) the corresponding model is
within the reach of the corresponding dark matter search
experiment.}\label{fig:dm_DELTAMH_1}
}
\FIGURE{\epsfig{file=dm_DELTAMH_2.eps,width=16cm}
\caption{Same as in Fig.\ref{fig:dm_DELTAMH_1}, but at $m_0=1450$
GeV.}\label{fig:dm_DELTAMH_2}
}

The overall summary and comparison of direct and indirect dark matter
detection methods considered here is shown in
Fig.~\ref{fig:dm_DELTAMH_1} 
($m_0=300$ GeV) and Fig.~\ref{fig:dm_DELTAMH_2} ($m_0=1450$
GeV) for the conservative Bukhert profile. 
This conservative assumption is why,
in both cases, the gamma ray flux from the Galactic Center is well below the
projected GLAST sensitivity. 
Antimatter searches will instead produce
positive results in the higgsino-like region at negative $\Delta
m_H$. In particular, antideuteron searches on GAPS look extremely
promising in the present setup, as they will entirely probe the
WMAP-allowed region for both values of $m_0$. In the $m_0=1450$ GeV
scenario (Fig.~\ref{fig:dm_DELTAMH_2}), as for the muon flux, antimatter
fluxes and gamma ray flux undergo a sudden drop motivated by the
gauge boson threshold ($m_{\widetilde{Z_1}}<m_W$). Light sfermions help
to marginally increase the antimatter rates (see right panels), but
the resulting fluxes are found to lie, in all cases, below the
expected experimental sensitivity.

\subsection{NUHM2 model: Collider searches for SUSY}

In this section, we illustrate aspects of the NUHM2 model that lead to
collider search possibilities not possible in the NUHM1 model. We
illustrate our points by showing in Table \ref{tab:2} three cases where
the NUHM2 model can be in accord with the WMAP measurement, but which
give rise to unique collider phenomenology.  All parameter points 
in this table have the same
values of $m_{1/2}=300$ GeV, $A_0=0$, $\tan\beta =10$, $\mu >0$ and
$m_t=178$ GeV.  For the first point, labeled NUHM2a, we take $m_0=300$
GeV, with $\mu =220$ GeV and $m_A =140$ GeV. It occurs in the lower-left
region of the plot in Fig.~\ref{fig:nuhm2_chi2} and gives a relic
density $\Omega_{\tz_1}h^2=0.10$. It is characterized by {\it both} a
low $\mu$ and a low $m_A$ value, unlike the NUHM1 model, which must have
one or the other small, but not both, to be in accord with WMAP. This
point yields light higgsinos {\it and} light $A$, $H$ and $H^\pm$ SUSY Higgs
bosons. The second point, NUHM2b, has $m_0=300$ GeV as well, but has
input parameters $m_{H_d}^2=(1651.7\ {\rm GeV})^2$ and
$m_{H_u}^2=-(1051.7\ {\rm GeV})^2$. It is characterized by relatively
light left-handed sleptons and sneutrinos, due to the large $S$ term in
the RGEs. Finally, we show NUHM2c, which has $m_0=1450$ GeV with
$m_{H_d}^2=(7047.3\ {\rm GeV})^2$ and $m_{H_u}^2=-(4147.3\ {\rm GeV})^2$. 
It is characterized by the presence of very light $\tu_R$ and $\tc_R$
squarks. Notice that the right sleptons are heavier than all the
squarks, and that $\ttau_1 \simeq \ttau_L$ is the lightest of the
charged sleptons\footnote{ 
There might also appear to be a possibility of
generating characteristic spectra with $S$ large and positive at the
GUT scale, where only $\tell_R$  are light, while
gluinos, charginos and neutralinos as well as most
squarks and left sleptons would be heavy, making the signal difficult to
detect at the LHC.
This case does not, however, seem to be
possible because large positive $S$ leads to $m_A^2<0$ and thus a 
breakdown in  the REWSB mechanism before the $\te_R$ becomes light enough.}.

\TABLE{
\begin{tabular}{lccc}
\hline
parameter & NUHM2a & NUHM2b & NUHM2c \\
\hline
$m_0$ & {\bf 300} & {\bf 300} & {\bf 1450} \\
$\mu$ & {\bf 220} & 933.2 & 3443.7 \\
$m_A$ & {\bf 140} & 1884.6 & 7765.1 \\
$m_{H_d}^2$ & $-(506.4)^2$ & ${\bf (1651.7)^2}$ & ${\bf (7047.3)^2}$ \\
$m_{H_u}^2$ & $-(263.5)^2$ & $-{\bf (1051.7)^2}$& $-{\bf (4147.3)^2}$ \\
$m_{\tg}$ & 726.4 & 739.4 & 807.8 \\
$m_{\tu_L}$ & 720.6 & 740.4 & 1724.8 \\
$m_{\tu_R}$ & 713.3 & 591.9 & 151.6 \\
$m_{\tst_1}$ & 491.0 & 661.9 & 1802.9 \\
$m_{\tb_1}$ & 629.0 & 730.6 & 1830.5 \\
$m_{\te_L}$ & 377.6 & 180.9 & 660.7 \\
$m_{\te_R}$ & 292.4 & 546.3 & 2316.1 \\
$m_{\ttau_1}$ & 290.1 & 149.3 & 522.9 \\
$m_{\tnu_\tau}$ & 368.3 & 129.9 & 513.1 \\
$m_{\tw_2}$ & 293.8 & 937.1 & 3428.8 \\
$m_{\tw_1}$ & 174.4 & 236.0 & 250.4 \\
$m_{\tz_4}$ & 296.4 & 935.7 & 3427.1 \\
$m_{\tz_3}$ & 228.5 & 931.1 & 3426.5 \\ 
$m_{\tz_2}$ & 178.7 & 236.1 & 251.5 \\ 
$m_{\tz_1}$ & 108.9 & 119.2 & 122.0 \\ 
$m_{H^+}$ & 162.1 & 1898.6 & 7816.6 \\
$m_h$ & 113.3 & 116.5& 120.3 \\
$\Omega_{\tz_1}h^2$& 0.10 & 0.17 & 0.14 \\
$BF(b\to s\gamma)$ & $4.2\times 10^{-4}$ & $3.3\times 10^{-4}$ & $3.5\times 10^{-4}$ \\
$\Delta a_\mu    $ & $15.3 \times  10^{-10}$ & $13.0\times 10^{-10}$ & $1.2\times 10^{-10}$ \\
\hline
\end{tabular}
\caption{\label{tab:2} Masses and parameters in~GeV units
for three NUHM2 models, where
$m_{1/2} =300$ GeV, $A_0=0$, $\tan\beta =10$ and $m_t=178$ GeV.
Input parameters are
shown as bold-faced.
}
\label{tab:nuhm2}
}
%

\subsubsection{Fermilab Tevatron}

The point NUHM2a with low $\mu $ and low $m_A$ values will be difficult
to probe at the Fermilab Tevatron. In this case, the chargino and
neutralino masses are still rather large, and $\tz_2\to\tz_1 e^+e^-$ has
a branching fraction of just 1.5\%, so that
trilepton signals will be difficult to detect above background. The
$m_A$ and $\tan\beta$ values are such that this point lies just beyond a
``hole'' in parameter space where none of the Higgs bosons are
accessible to the Tevatron, even at the 95\%CL exclusion
level~\cite{bht,carena}.  The point NUHM2b, with relatively light
sleptons, will also be difficult to probe at the Tevatron since charginos
are quite heavy and slepton pairs are difficult to detect for any mass
choices~\cite{bcpt_sl}. The point NUHM2c has {\it two} flavors 
of relatively light
squarks -- $\tu_R$ and $\tc_R$ -- but the squark-neutralino mass gap is
rather small. The signal would be identical to the one searched for in
Ref.~\cite{cdfstop}, except with essentially twice the expected cross section
for any given squark and LSP masses since 
$\sigma(\tc_R\tc_R)\simeq \sigma(\tu_R\tu_R)$). 
It is possible that a dedicated squark search might
be able to detect a signal in the dijet$+\eslt$ channel, where relatively
low jet $E_T$ and $\eslt$ values $\sim 25-50$ GeV might be expected,
owing to the small $\tu_R -\tz_1$ mass gap. Alternatively, if it becomes
possible to tag $c$-jets with significant efficiency, it may be possible 
to suppress backgrounds sufficiently to pull out the signal. 
The phenomenology of light $\tu_R$ and $\tc_R$ squarks for the Tevatron 
is discussed more completely in Ref. \cite{auto2}.


\subsubsection{CERN LHC}

Squarks and gluinos would be produced at large rates at the CERN LHC. 
Their cascade decays would, in general, lead to the production of Higgs
bosons in SUSY events. In the NUHM2 model, since the Higgs sector is
essentially arbitrary, the heavier Higgs bosons $H, A$ and $H^{\pm}$
could decay into other sparticles, resulting in characteristic events at
the LHC\cite{higgsdk2susy}. 
In the case of the NUHM2a scenario shown in the table, the low value of $\mu$
implies the entire spectrum of charginos and neutralinos will be quite
light, and accessible via squark and gluino cascade decays. The cascade
decays will be much more complex than in a typical mSUGRA scenario, but
as for the NUHM1b point discussed earlier, potentially offer a rich
possibility for extracting information via a variety of mass edges that
would be theoretically present. The feasibility of actually doing so
would necessitate detailed simulations beyond the scope of the present
analysis.
In addition, in this scenario, the heavier Higgs
bosons are also quite light, and in fact the heavier inos are able to
decay into $A$, $H$ and $H^\pm$ with significant rates. The production
of $H$ and $A$ followed by $H,\ A\to \tau^+\tau^-$ and perhaps also 
to $\mu^+\mu^-$, should be detectable at the LHC. In addition,
$Wh$ or $t\bar{t}h$ production with $W$ and one of the tops decaying
leptonically should also be detectable.

The point NUHM2b is characterized by light sleptons. In this case,
$\tq_L$ cascade decays will be lepton-rich, since $\tz_2\to
\te_L e$ with a branching ratio $\sim 10\%$, while
$\tw_1 \to \tell_L\nu+\ell\tnu$ essentially 100\% of the time.  The
decay $\tz_2\to\tnu \nu$ is also allowed, but in this case the sneutrino
decays invisibly.  Gluinos nearly always decay to $\tu_R u$
or $\tc_R c$. The squarks then decay via $\tu_R\to u\tz_1$ and $\tc_R\to
c\tz_1$, so that $\tg\tg$ production will give rise to $\sim 4$ jet
$+\eslt$ events.

The point NUHM2c, characterized by light $\tu_R$ and $\tc_R$ squarks,
will allow for squark production at very large rates. In addition,
$\tg\tg$ production followed by $\tg\to \tu_R u$ or $\tc_R c$ will give
rise to $2-4$-jet $+\eslt$ events, since the jets from $\tu_R\to u\tz_1$
decay will be rather soft owing to the small $\tu_R -\tz_1$ mass
gap. The cascade decay events in this case will be lepton-poor since
$\tz_2$ decays mostly to $\tz_1 h$ or $\tu_R u,\ \tc_R c$ final states,
and $\tq_L$ is heavy so that charginos are not abundantly produced via
their decays.

\subsubsection{Linear $e^+e^-$ collider}

Any scenario similar to that represented by point NUHM2a would be a
bonanza for the ILC. In this case, a $\sqrt{s}=0.5-0.6$ TeV machine
would be able to access both chargino and all four neutralino states as
well as the heavy Higgs bosons $H$, $A$ and $H^\pm$. The difficulty
would be in sorting out the large number of competing reactions, but
here, variable center of mass energy and beam polarization would be a
huge help.  A complete reconstruction of chargino and neutralino mass
matrices may be possible~\cite{zerw}. The low value of $m_A$ should serve
to distinguish this case from a NUHM1a-like scenario.

In the case of NUHM2b, the $\ttau_1$ and $\te_L$ slepton
states would be accessible to early searches, along with
$\tw_1^+\tw_1^-$ production. Beam polarization would be a key ingredient
in determining that the $\ttau_1$ and $\te_L$ are left-handed.  
Determination that $\ttau_1$ is dominantly $\ttau_L$ and/or $m_{\te_L}
\ll m_{\te_R}$ would already point to an unconventional scenario. 

The case of NUHM2c would allow $\tu_R\bar{\tu}_R$ and $\tc_R\bar{\tc}_R$
production to occur at large rates at an ILC. Again, the beam
polarization would easily determine the right-hand nature of these
squarks, which would be a key measurement. It should also be possible to
determine their masses~\cite{FF}, and if it is possible to tag $c$ jets
with reasonable efficiency, to also distinguish between squark flavors. 

We note here that in addition, in the NUHM2 model, the reach of an ILC
may be far greater than the CERN LHC for supersymmetry. The reason is
that the LHC reach is mainly determined by the $m_0$ and $m_{1/2}$
parameters, which determine the overall squark and gluino mass scales.
In contrast, the ILC reach for chargino pair production depends strongly
on the $\mu$ parameter. Thus, the NUHM2 case where $m_0$ and $m_{1/2}$
are large, while $\mu$ is small may mean chargino pair production is
accessible to an ILC while gluino and squark pair production is beyond
LHC reach. The case is illustrated in Fig.~\ref{fig:nuhm2_col}, where we
show the $m_0\ vs.\ m_{1/2}$ plane for $A_0=0$, $\tan\beta =10$, $\mu
>0$ $m_t=178$ GeV and {\it a}) $\mu =m_A=500$ GeV and {\it b}) $\mu =
m_A= 300$ GeV.  The yellow and green regions are WMAP allowed, while the
unshaded regions have $\Omega_{\tz_1}h^2$ bigger than the WMAP upper
bound. The yellow bands just above the LEP excluded blue regions in both
frames is where $2m_{\tz_1} \simeq m_h$. The corresponding band in the
left panel at $m_{1/2}\simeq 0.6$~TeV is the $A$ funnel, while in the upper
yellow/green regions in both panels the LSP has a significant higgsino
content. 

The SUSY reach of the CERN LHC should be similar to the case of the
mSUGRA model calculated in Ref.~\cite{lhcreach}, and as before, we show
this result as an approximate depiction of the LHC reach for the case
for the NUHM2 model.  We also show the mass contour in {\it a}) for a
250 and 500 GeV chargino, accessible to a $\sqrt{s}=0.5$ or 1 TeV ILC
machine. Here, the 1 TeV machine has a reach beyond the large $m_0$
reach of the LHC. In the case of frame {\it b}), $\mu$ is so small that
the $\tw_1$ mass is almost always below $\sim 330$ GeV, and so the
entire plane shown would be accessible to a 1 TeV ILC!

\FIGURE{\epsfig{file=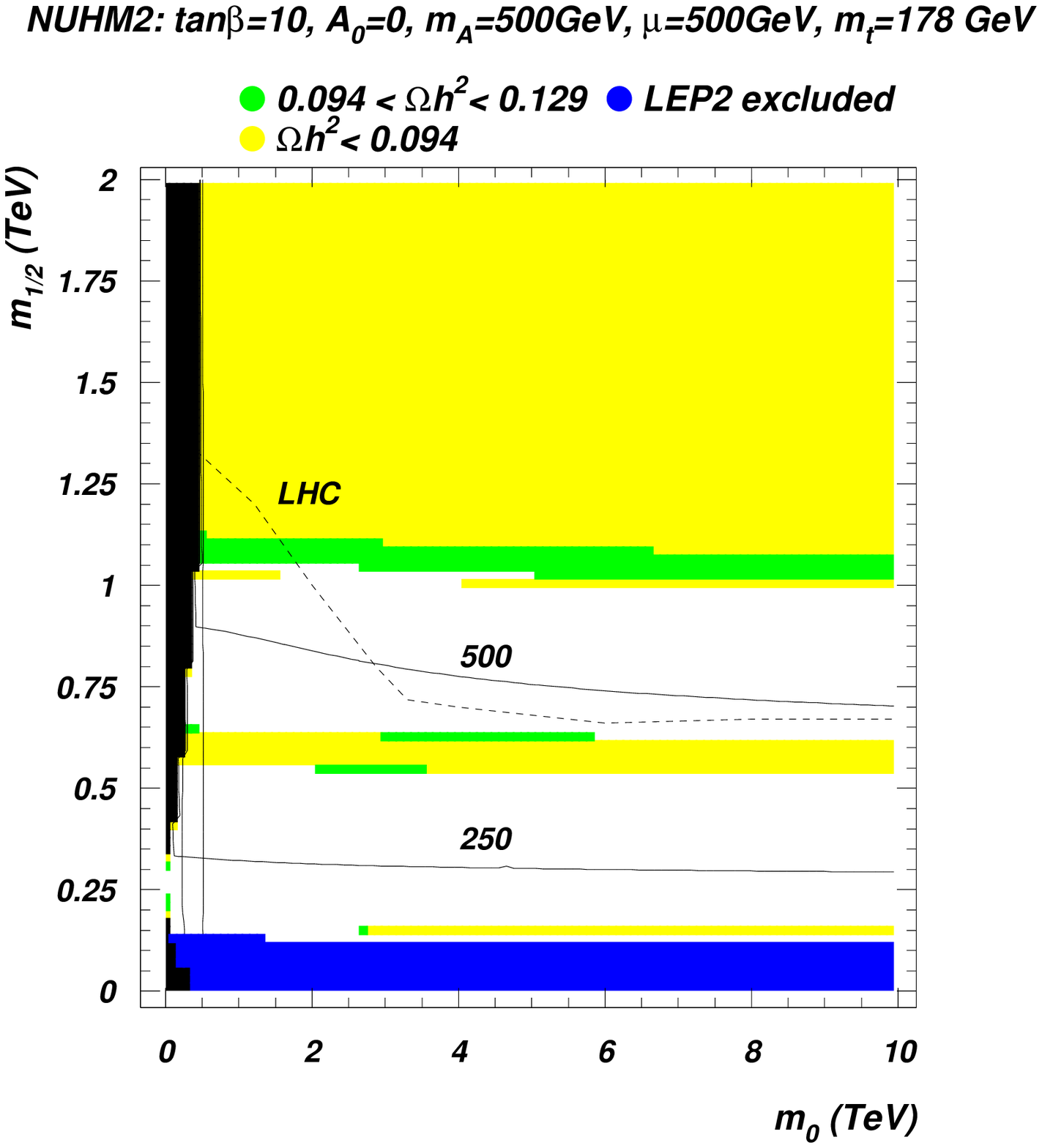,height=8cm}
\epsfig{file=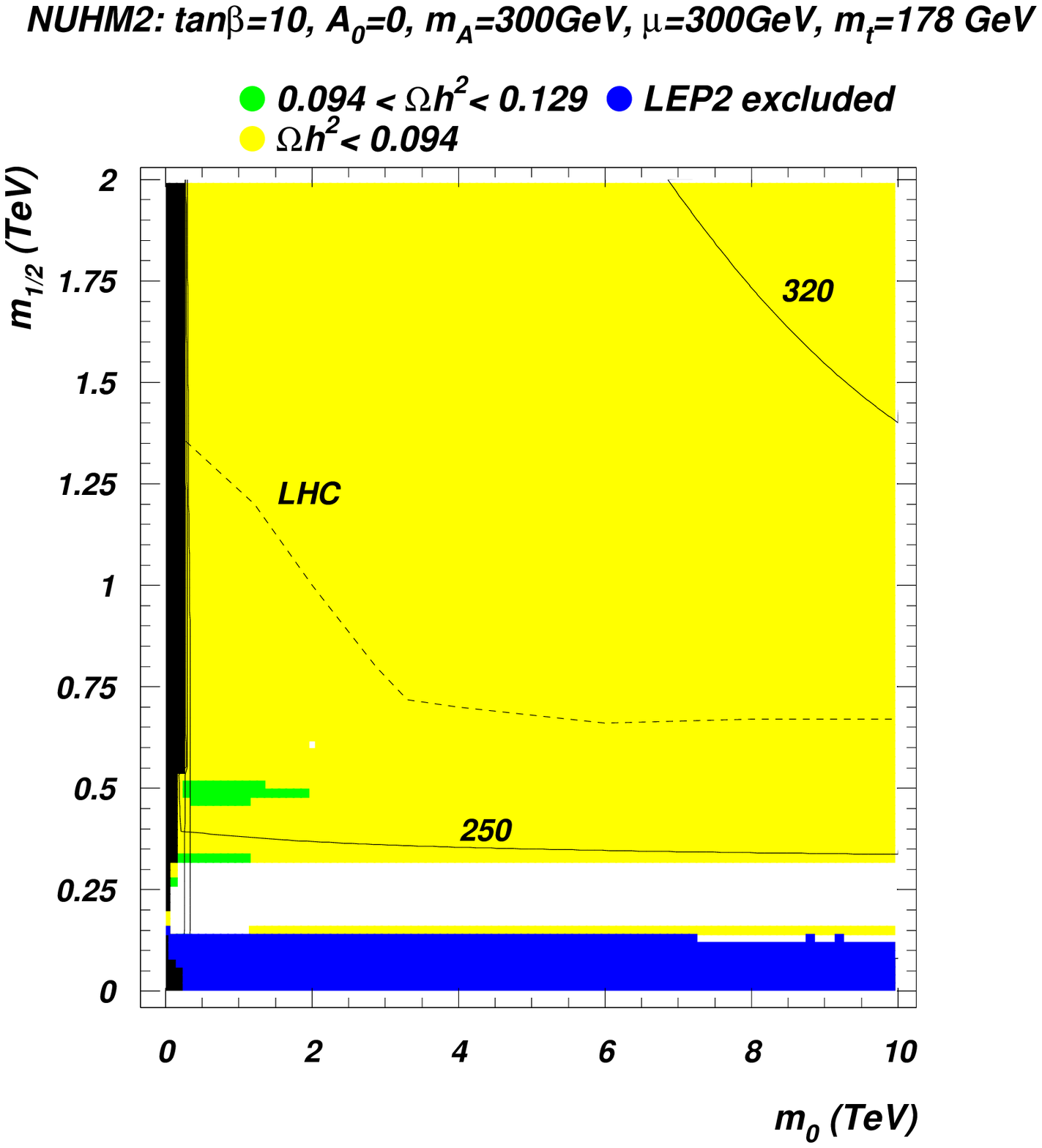,height=8cm}
\caption{Approximate reach of the CERN LHC (100 fb$^{-1}$) and
ILC for supersymmetric matter in the NUHM2 model in the
$m_0\ vs.\ m_{1/2}$ plane for $\tan\beta=10$, $A_0=0$, 
$\mu>0$ and $m_t=178$ GeV, for {\it a}) $\mu =500$ GeV, $m_A=500$ GeV
and {\it b}) $\mu =300$ GeV, $m_A=300$ GeV.}
\label{fig:nuhm2_col}
}

\section{Concluding Remarks}\label{sec:conclusions}

We have examined the  phenomenological implications of 
gravity-mediated SUSY breaking models with universal matter scalars, 
but with non-universal Higgs soft SUSY breaking masses. 
For simplicity, we assume a common GUT scale mass parameter for {\it
all} matter scalars -- this guarantees that phenomenological
constraints from flavor
physics are respected -- but unlike in mSUGRA, entertain the possibility
that the soft SUSY breaking mass parameters in the Higgs sector are
unrelated to the matter scalar mass. In these non-universal Higgs mass
(NUHM) models where the Higgs fields $H_u$ and $H_d$ 
originate in a common multiplet (as, for instance, in an $SO(10)$ model
with a single Higgs field), we would have $m_{H_u}^2= m_{H_d}^2$, and
there would be just one additional new parameter (NUHM1
model)~\cite{bbmpt} {\it vis \`a vis} mSUGRA, while the more general
scenario would have two additional parameters (NUHM2 model).
We have found that, once WMAP constraints are
incorporated into the analysis, this seemingly innocuous extension of
the mSUGRA parameter space, which naively would not be expected to
affect squark, gluino and slepton masses, significantly expands the
possibilities for LHC and linear collider phenomenology from mSUGRA
expectations: phenomena that were considered unlikely because they were
expected to occur only in particular corners of parameter space become
mainstream in the extended model. In the absence of any compelling
theory of sparticle masses, the necessity for ensuring that all
experimental possibilities are covered is sufficient reason to examine
the consequences of NUHM models, regardless of whether or not one
considers these to be theoretically attractive.

We have examined the allowed parameter space of the NUHM models with respect
to neutralino relic density (WMAP), $BF(b\to s\gamma )$ and $\Delta a_\mu$
constraints. 
The WMAP upper bound on $\Omega_{CDM}h^2$ requires 
rather efficient 
LSP annihilation, and so severely restricts
any supersymmetric model. 
Aside from the bulk region with small values of bino and scalar masses,
enhanced LSP annihilation may occur if the LSP has significant higgsino
or wino components (the latter is not possible in models with unified
gaugino masses\footnote{For an exception to this, see the {\bf 200}
model in Ref.\cite{nonuni}.}), resonantly annihilates via 
Higgs scalars (or $Z$ bosons),
or co-annihilates with the stau or some other charged sparticle. Within
mSUGRA, the higgsino annihilation region occurs only if $m_0$ is very
large, while resonance annihilation with heavy Higgs scalars is possible
only for large values of $\tan\beta$.  However, even in the simple one
parameter NUHM1 extension of mSUGRA, for almost any values of $m_0$,
$m_{1/2}$ and $\tan\beta$, there are two different choices of $m_\phi$
(defined in the text) that can bring the relic density to be in accord
with the WMAP measurement: for large positive $m_\phi$, one enters the
higgsino region, while for large negative values of $m_\phi$, one enters
the $A$ annihilation funnel.  The higgsino region with small $\mu$
values gives rise to large rates for direct and indirect detection of
neutralino dark matter, and also leads to light charginos and
neutralinos which might be accessible to a TeV-scale linear $e^+e^-$
collider, or which can enrich the gluino and squark cascade decays
expected at the CERN LHC.  The $A$-funnel region in the NUHM1 model can
also occur at any $\tan\beta$ value, and usually
leads to relatively light $H,\ A$ and $H^\pm$ Higgs bosons which may be
accessible to collider searches.  Also, the expected suppression of $e$
and $\mu$ signals from cascade decays, which is expected in the mSUGRA
model for points in the $A$-funnel due to enhanced -ino decays to
taus\cite{ltanb}, will not necessarily obtain in the NUHM1 model since
$\tan\beta$ is not required to be large.  Since the early universe
neutralino pair annihilation cross sections are enhanced on the
$A$-resonance, indirect DM signals are, in general, enhanced as well.

The parameter freedom is enhanced even more in the NUHM2 model. In this
case, the mSUGRA-fixed parameters $\mu$ and $m_A$ can now be taken as
inputs, rather than outputs. This allows one to always dial in a low
value of $\mu$ such that one is in the higgsino region, or a low value
of $m_A$ so that one is in the $A$-funnel. As before, direct and
indirect DM detection rates are enhanced in these regions. Collider
signals may change as well, since now {\it all} charginos and
neutralinos can be light, and
one can have enhanced cascade
decays of squarks and gluinos to charginos, neutralinos {\it and} to
heavy Higgs bosons. In the case where $m_0$ and $m_{1/2}$ are large, but
$\mu$ is small, the reach of a LC may exceed that of the CERN LHC. In
the NUHM2 model, qualitatively new regions emerge where the relic
density is suppressed due to novel sparticle mass patterns: very light
left-handed sleptons, or very light right-handed up and charm squarks,
which have obvious implications for collider signals.  The latter case
results in large rates for direct and indirect detection of neutralino
DM, in addition to large $jet+\eslt$ signals at hadron colliders, and to
the possibility of squark NLSPs at $e^+e^-$ LCs.

In conclusion, 
we have seen that the seemingly
innocuous decoupling of scalar Higgs mass parameters from other scalar
masses can significantly alter our expectations of what we may expect 
in terms of dark matter as well as (s)particle physics phenomenology.
This is partly because altering the Higgs potential can dramatically
change the value of $\mu^2$ that yields the correct value of $M_Z^2$,
and partly because renormalization group evolution of sparticle mass
parameters is dramatically
altered by a non-zero value of $S$ in the NUHM2 model.  
For
both the NUHM1 and NUHM2 extensions of the mSUGRA model
one is  able to
find generic regions of parameter space that are in good agreement 
with the WMAP determination of the cold dark matter relic density,
as well as with constraints from $b\to s\gamma$ and $(g-2)_{\mu}$.
These regions can lead to distinctive
signals at both direct and indirect dark matter detection experiments,
and also provide distinctive signatures at both 
the CERN LHC $pp$ collider and the International Linear Collider, with 
a center of mass energy $\sqrt{s}=0.5-1$ TeV.

\acknowledgments

This research was supported in part by grants from
the U.S. Department of Energy.

%

\end{document}